\let\accentvec\vec
\documentclass[runningheads,a4paper]{llncs}

\let\vec\accentvec
\usepackage{amsmath}								
\usepackage{amssymb}
\usepackage[misc,geometry]{ifsym}    
\usepackage{makeidx}  
\usepackage{lipsum}
\usepackage{lastpage}
\usepackage{clrscode}
\usepackage{bm} 
\usepackage{multirow}
\usepackage{array}
\usepackage{graphicx}
\usepackage{picinpar}
\usepackage{enumerate}
\usepackage{epsfig}
\usepackage{boxedminipage}
\usepackage{indentfirst}
\usepackage{afterpage}
\usepackage{yhmath}
\usepackage{graphicx}	
\usepackage{diagbox}
\usepackage{extarrows}
\usepackage{color}
\usepackage{float}
\usepackage[colorlinks, linkcolor=blue, anchorcolor=blue, citecolor=blue]{hyperref}
\usepackage{setspace}
\usepackage[dvipsnames]{xcolor}
\usepackage{algorithm,algorithmic}
\usepackage{booktabs}
\usepackage{longtable}
\usepackage{tabu}
\def\sign{{\rm sign}}

\begin{document}

\title{GSplit LBI: Taming the Procedural Bias in Neuroimaging for Disease Prediction}
\titlerunning{GSplit LBI: Taming the Procedural Bias in Neuroimaging}  

\author{Xinwei Sun\inst{1} \and Lingjing Hu\inst{2}(\Letter)  \and Yuan Yao\inst{3}(\Letter) \and Yizhou Wang\inst{4}} 
\institute{School of Mathematical Science, Peking University, Beijing, 100871, China \and Yanjing Medical College, Capital Medical University, Beijing, 101300, China \and Hong Kong University of Science and Technology and Peking University, China \and National Engineering Laboratory for Video Technology, Key Laboratory of Machine Perception, School of EECS, Peking University, Beijing, 100871, China  }
\authorrunning{Xinwei Sun et al.} 
\maketitle              
\begin{abstract}
In voxel-based neuroimage analysis, lesion features have been the main focus in disease prediction due to their interpretability with respect to the related diseases. However, we observe that there exist another type of features introduced during the preprocessing steps and we call them ``\textbf{Procedural Bias}". Besides, such bias can be leveraged to improve classification accuracy. Nevertheless, most existing models suffer from either under-fit without considering procedural bias or poor interpretability without differentiating such bias from lesion ones. In this paper, a novel dual-task algorithm namely \emph{GSplit LBI} is proposed to resolve this problem. By introducing an augmented variable enforced to be structural sparsity with a variable splitting term, the estimators for prediction and selecting lesion features can be optimized separately and mutually monitored by each other following an iterative scheme. Empirical experiments have been evaluated on the Alzheimer's Disease Neuroimaging Initiative\thinspace(ADNI) database. The advantage of proposed model is verified by improved stability of selected lesion features and better classification results.  
\end{abstract}

\section{Introduction}
\numberwithin{equation}{section}

\begin{figure}
\centering
\begin{minipage}[t]{0.22\linewidth}
    \includegraphics[width= \columnwidth]{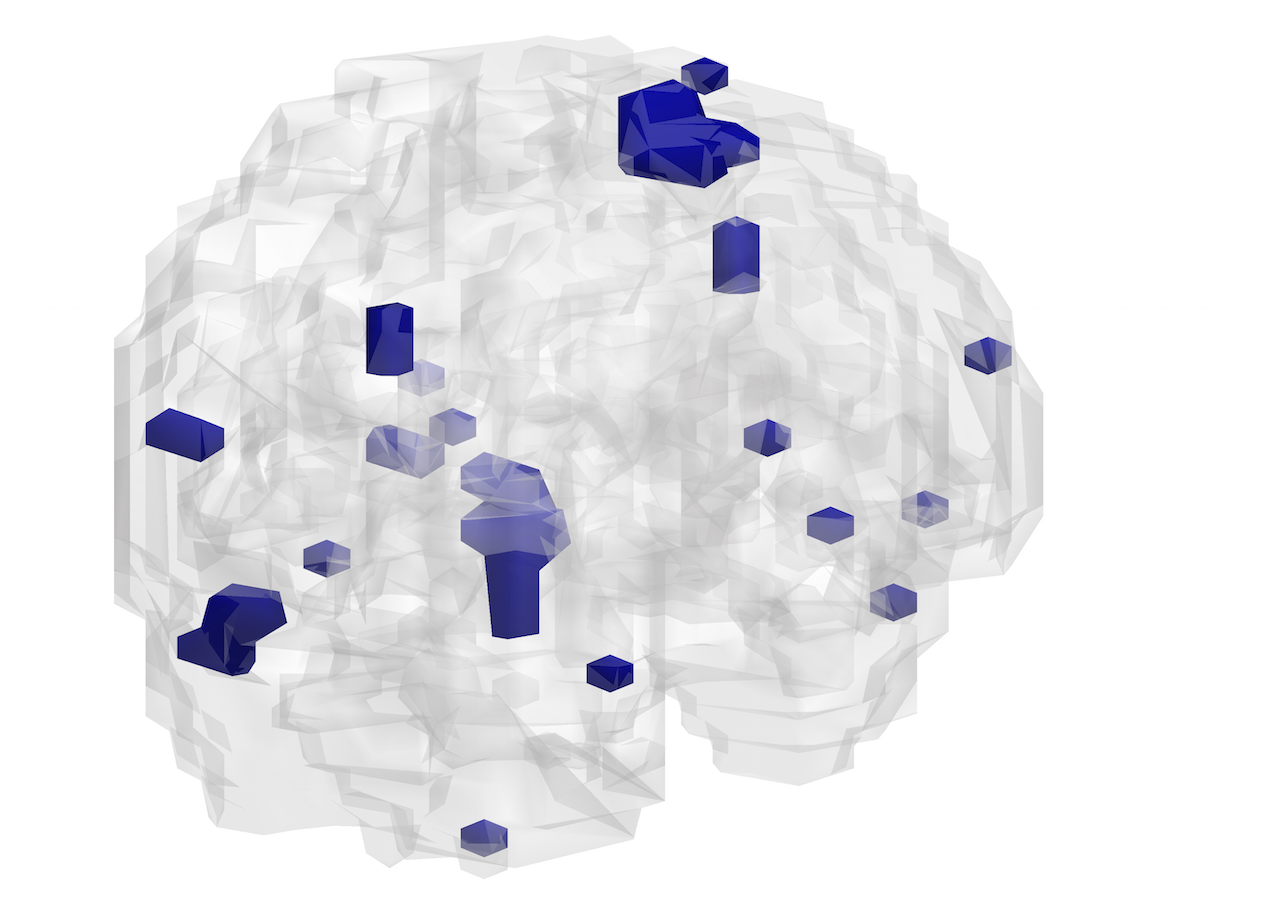}
    \end{minipage}
\begin{minipage}[t]{0.18\linewidth}
    \includegraphics[width= \columnwidth]{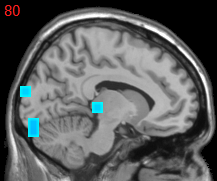}
\end{minipage}
\begin{minipage}[t]{0.1501\linewidth}
    \includegraphics[width= \columnwidth]{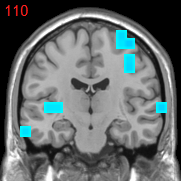}
\end{minipage}
\begin{minipage}[t]{0.1252\linewidth}
    \includegraphics[width= \columnwidth]{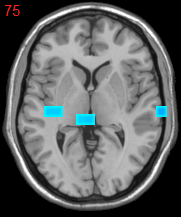}
\end{minipage}
\caption{The overlapped voxels among top 150 negative value voxels in each fold of $\beta_{pre}$ at the time corresponding to the best average prediction result in the path of GSplit LBI using 10-fold cross-validation. For subjects with AD, they represent enlarged GM voxels surrounding lateral ventricle, subarachnoid space, edge of gyrus, etc.}\label{figure:7}
\end{figure}

Usually, the first step of voxel-based neuroimage analysis requires preprocessing the T$_{1}$-weighted image, such as segmentation and registration of grey matter\thinspace(GM), white matter\thinspace(WM) and cerebral spinal fluid\thinspace(CSF). However, some systematic biases due to scanner difference and different population etc., can be introduced in this pipeline \cite{whyvoxel-based}. Part of them can be helpful to the discrimination of subjects from normal controls\thinspace(NC), but may not be directly related to the disease.
For example in structural Magnetic Resonance Imaging\thinspace(sMRI) images of subjects with Alzheimer's Disease\thinspace(AD), after spatial normalization during simultaneous registration of GM, WM and CSF, the GM voxels surrounding lateral ventricle and subarachnoid space etc. may be mistakenly enlarged caused by the enlargement of CSF space in those locations \cite{whyvoxel-based} compared to normal template, as shown in Fig.~\ref{figure:7}. 
Although these voxels/features are highly correlated with disease, they can't be regarded as lesion features in an interpretable model. In this paper we refer to them as ``\textbf{Procedural Bias}", which should be identified but is neglected in the literature. We observe that it can be harnessed in our voxel-based image analysis to improve the prediction of disease. 

Together with procedural bias, the lesion features are vital for prediction and lesion regions analysis tasks, which are commonly solved by two types of regularization models. Specifically, one kind of models such as general losses with $l_{2}$ penalty, elastic net \cite{elasticnet} and graphnet \cite{graphnet} select strongly correlated features to minimize classification error. However, such models don't differentiate features either introduced by disease or procedural bias and may also introduce redundant features. Hence, the interpretability of such models are poor and the models are prone to over-fit. The other kind of models with sparsity enforcement such as TV-$L_{1}$\thinspace(Combination of Total Variation \cite{TV} and $L_{1}$) and particularly $n^{2}$ GFL \cite{n2gfl} enforce strong prior of disease on the parameters of the models introduced in order to capture the lesion features. Although such features are disease-relevant and the selection is stable, the models ignore the inevitable procedural bias, hence, they are losing some prediction power.

To incorporate both tasks of prediction and selection of lesion features, we propose an iterative dual-task algorithm namely \emph{Generalized Split LBI}\thinspace(GSplit LBI) which can have better model selection consistency than generalized lasso \cite{Genlasso}. Specifically, by the introduction of variable splitting term inspired by Split LBI \cite{Splitlbi}, two estimators are introduced and split apart. One estimator is for prediction and the other is for selecting lesion features, both of which can be pursued separately with a gap control. Following an iterative scheme, they will be mutually monitored by each other: the estimator for selecting lesion features is gradually monitored to pursue stable lesion features; on the other hand, the estimator for prediction is also monitored to exploit both the procedural bias and lesion features to improve prediction. To show the validity of the proposed method, we successfully apply our model to voxel-based sMRI analysis for AD, which is challenging and attracts increasing attention. 

\section{Method}

\subsection{GSplit LBI Algorithm}
Our dataset consists of $N$ samples $\{x_{i},y_{i}\}_{1}^{N}$ where $x_{i} \in \mathbb{R}^{p}$ collects the $i^{th}$ neuroimaging data with $p$ voxels and $y_{i} = \{\pm 1\}$ indicates the disease status ($-1$ for Alzheimer's disease in this paper). 
$X \in \mathbb{R}^{N \times p}$ and $y \in \mathbb{R}^{p}$ are concatenations of $\{x_{i}\}_{i}$ and $\{y_{i}\}_{i}$. Consider a general linear model to predict the disease status (with the intercept parameter $\beta_{0} \in \mathbb{R}$),
\begin{equation}
\log P( y_i=1| x_i) - \log P(y_i=-1|x_i) =  x_i^T \beta_{pre} + \beta_{0}. 
\end{equation} 
A desired estimator $\beta_{pre} \in \mathbb{R}^p$ should not only fit the data by maximizing the log-likelihood in logistic regression, but also satisfy the following types of structural sparsity: (1) the number of voxels involved in the disease prediction is small, so $\beta_{pre}$ is sparse; (2) the voxel activities should be geometrically clustered or 3D-smooth, suggesting a TV-type sparsity on $D_G \beta_{pre}$ where $D_{G}$ is a graph difference operator\footnote{Here $D_G:\mathbb{R}^V \to \mathbb{R}^E$ denotes a graph difference operator on $G=(V,E)$, where $V$ is the node set of voxels, $E$ is the edge set of voxel pairs in neighbour (e.g. 3-by-3-by-3), such that $D_G(\beta)(i,j):=\beta(i)-\beta(j)$.}; (3) the degenerate GM voxels in AD are captured by nonnegative component in $\beta_{pre}$. However, the existing procedural bias may violate these \emph{a priori} sparsity properties, \emph{esp.} the third one, yet increase the prediction power. 

To overcome this issue, we adopt a variable splitting idea in \cite{Splitlbi} by introducing an auxiliary variable $\gamma \in \mathbb{R}^{|V|+|E|}$ to achieve these sparsity requirements separately, while controlling the gap from $D\beta_{pre}$ with penalty $S_{\rho}(\beta_{pre},\gamma) := \Vert D\beta_{pre} - \gamma \Vert_{2}^{2} := \Vert \beta_{pre} - \gamma_{V} \Vert_{2}^{2}  + \Vert \rho D_{G}\beta_{pre} - \gamma_{G} \Vert_{2}^{2}$ with $\displaystyle \gamma = \left[ \begin{array}{cc} \gamma_{V}^T & \gamma_{G}^T \end{array} \right]^T$ and $\displaystyle D = \left[ \begin{array}{cc} I & \rho D_{G}^T\end{array} \right]^T$.
Here $\rho$ controls the trade-off between different types of sparsity. Our purpose is thus of two-folds: (1) use $\beta_{pre}$ for prediction; (2) enforce sparsity on $\gamma$. Such a dual-task scheme can be illustrated by Fig.~\ref{figure:2}. 
\begin{figure}[t]
\centering
    \includegraphics[width= 0.991\columnwidth]{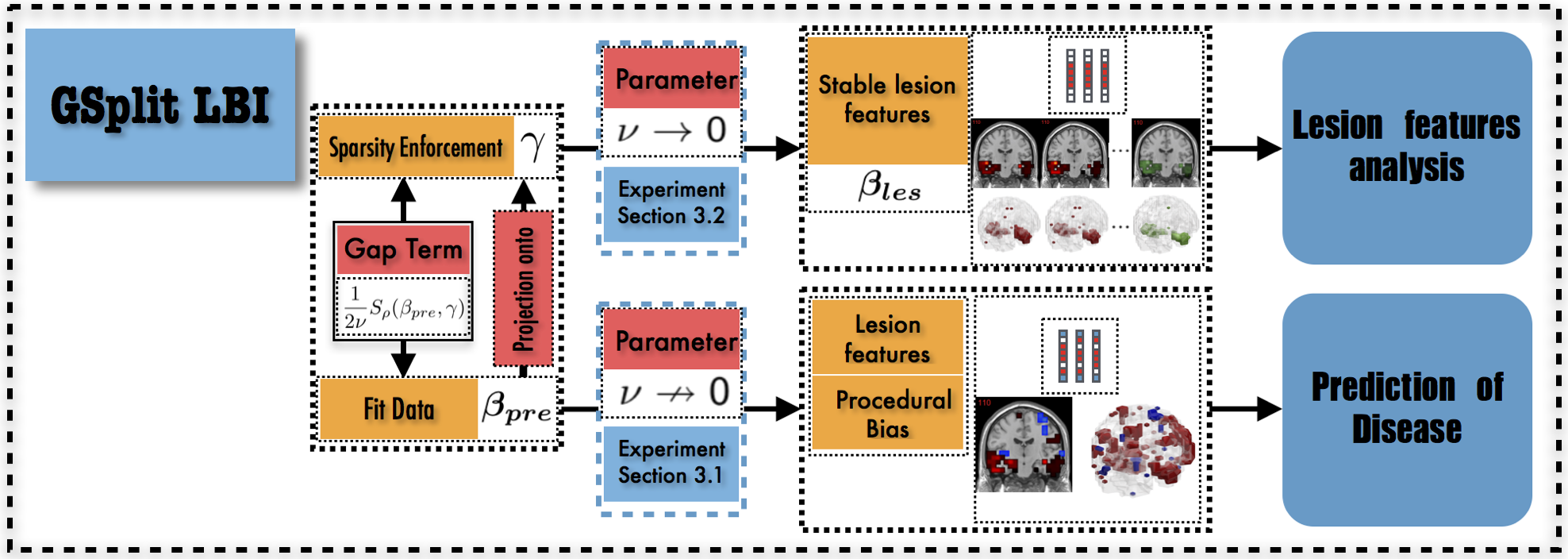}
\caption{Illustration of \emph{GSplit LBI}. The gap between $\beta_{pre}$ for fitting data and $\gamma$ for sparsity is controlled by $S_{\rho}(\beta_{pre},\gamma)$. The estimate $\beta_{les}$, as a projection of $\beta_{pre}$ on support set of $\gamma$, can be used for stable lesion features analysis when $\nu \to 0$ (Section~\ref{lesion}). When $\nu \nrightarrow 0$ (Section~\ref{prediction}) with appropriately large value, $\beta_{pre}$ can be used for prediction by capturing both lesion features and procedural bias.}
\label{figure:2}
\end{figure}

To implement it, we generalize the Split Linearized Bregman Iteration (Split LBI) algorithm in \cite{Splitlbi} to our setting with generalized linear models (GLM) and the three types of structural sparsity above, hence called Generalized Split LBI (or GSplit LBI). Algorithm~\ref{alg:1} describes the procedure with a new loss:
\begin{eqnarray}
\ell(\beta_{0},\beta_{pre},\gamma;\{x_{i},y_{i}\}_{1}^{N},\nu) := \ell(\beta_{0},\beta_{pre};\{x_{i},y_{i}\}_{1}^{N}) + \frac{1}{2\nu} S_\rho(\beta_{pre},\gamma), \label{2.1}
\end{eqnarray}
where $\ell(\beta_{pre};\{x_{i},y_{i}\}_{1}^{N})$ is the negative log-likelihood function for GLM and $\nu>0$ tunes the strength of gap control. The algorithm returns a sequence of estimates as a regularization path, $\{\beta_0^k, \beta_{pre}^{k},\gamma^{k},\beta_{les}^{k}\}_{k\geq 0}$. In particular, $\gamma^k$ shows a variety of sparsity levels and $\beta_{pre}^k$ is generically dense with different prediction powers. The projection of $\beta_{pre}^k$ onto the subspace with the same support of $\gamma^k$ gives estimate $\beta_{les}^k$, satisfying those \emph{a priori} sparsity properties (sparse, 3D-smooth, nonnegative) and hence being regarded as the interpretable lesion features for AD. The remainder of this projection is heavily influenced by procedural bias; in this paper the non-zero elements in $\beta_{pre}^k$ which are negative (-1 denotes disease label) with comparably large magnitude are identified as procedural bias, while others with tiny values can be treated as nuisance or weak features. In summary, $\beta_{les}$ only selects lesion features; while $\beta_{pre}$ also captures additional procedural bias. Hence, such two kinds of features can be differentiated, as illustrated in Fig.~\ref{figure:2}.
\begin{algorithm}[t]
\caption{GSplit LBI}
\label{alg:1}
	\begin{algorithmic}[1]
         \STATE  $\textbf{Input:}$ Loss function $\ell(\beta_{0},\beta_{pre},\gamma;\{x_i,y_i\}_{i=1}^{N},\nu)$, parameters $\nu$, $\rho$, $\kappa$, $\alpha>0$. 

        	\STATE $\textbf{Initialize:}$ $k = 0$, $t^{k} = 0$, $\beta_{0}^{k} = 0$, $\beta_{les}^{k} = 0$, $\beta_{pre}^{k} = 0$, $\gamma_{V}^{k} = 0_{p}$, $	\gamma_{G}^{k} = 0_{m}$, $z_{V}^{k} = 0_{p}$, $z_{G}^{k} = 0_{m}$ and $S_{k} := \mathrm{supp}(\gamma^{k}) = \emptyset$.

	\STATE $\textbf{Iteration}$  
        \STATE \quad    $\beta_{0}^{k+1} = \beta_{0}^{k} - \kappa \alpha \triangledown_{\beta_{0}} \thinspace \ell(\beta_{0}^{k}, \beta_{pre}^{k}, \gamma^{k};\{x_{i},y_{i}\}_{1}^{N},\nu)$

	\STATE \quad    $\beta_{pre}^{k+1} = \beta_{pre}^{k} - \kappa \alpha \triangledown_{\beta_{pre}} \thinspace \ell(\beta_{0}^{k}, \beta_{pre}^{k}, \gamma^{k};\{x_{i},y_{i}\}_{1}^{N},\nu)$ 

	\STATE \quad   $z^{k+1} = z^{k} - \alpha \triangledown_{\gamma} \thinspace \ell(\beta_{0}^{k}, \beta_{pre}^{k}, \gamma^{k};\{x_{i},y_{i}\}_{1}^{N},\nu) $  

	\STATE \quad	$\gamma_{V}^{k+1} = \kappa \cdot \mathcal{S}^{+}(z_{V}^{k+1},1)$, where $\mathcal{S}^{+}(x,1) = \max(x - 1,0)$ 
	
	\STATE \quad   $\gamma_{G}^{k+1} = \kappa \cdot \mathcal{S}(z_{G}^{k+1}, 1)$, where $\mathcal{S}(x, 1) = \sign(x) \cdot \max(|x| - 1,0)$ 
	                               
	\STATE \quad	$\beta_{les}^{k+1} = P_{S_{k+1}}\beta_{pre}^{k+1}$, where $P_{S} = P_{ker(D_{S^{c}})} =  I - D_{S^c}^{\dagger}D_{S^c}$

	\STATE \quad	$t^{k+1} = (k+1)\alpha$

         \STATE $\textbf{Output:}$ $\{\beta_0^k, \beta_{pre}^{k}, \beta_{les}^{k}, \gamma^{k}\}$,
	where $\gamma^{k+1} = \left[ \begin{array}{cc} \gamma_{V}^{k+1} \\ \gamma_{G}^{k+1} \end{array} \right]$ and $z^{k+1} = \left[ \begin{array}{cc} z_{V}^{k+1} \\ z_{G}^{k+1} \end{array} \right]$. 

	\end{algorithmic}
\end{algorithm}
\subsection{Setting the Parameters}

A stopping time at $t^{k}$ (line 10) is the regularization parameter, which can be determined via cross-validation {to minimize the prediction error \cite{bregman}. Parameter $\rho$ is a tradeoff between geometric clustering and voxel sparsity. Parameter $\kappa$, $\alpha$ is damping factor and step size, which should satisfy $\kappa \alpha \leq \nu / \kappa(1 + \nu \Lambda_{H} + \Lambda_{D}^{2})$ to ensure the stability of iterations. Here $\Lambda_{(\cdot)}$ denotes the largest singular value of a matrix and $H$ denotes the Hessian matrix of $\ell(\beta_{0},\beta_{pre};\{x_{i},y_{i}\}_{1}^{N})$}.

Parameter $\nu$ balances the prediction task and sparsity enforcement in feature selection. In this paper, it is task-dependent, as shown in Fig.~\ref{figure:2}. For prediction of disease, $\beta_{pre}$ with appropriately larger value of $\nu$ may increase the prediction power by harnessing both lesion features and procedural bias. For lesion features analysis, $\beta_{les}$ with a small value of $\nu$ is helpful to enhance stability of feature selection. For details please refer to supplementary information.

\section{Experimental Results}
We apply our model to AD/NC classification (namely ADNC) and MCI (Mild Cognitive Impairment)/NC (namely MCINC) classification, which are two fundamental challenges in diagnosis of AD. The data are obtained from ADNI\footnote{http://adni.loni.ucla.edu} database, which is split into 1.5T and 3.0T (namely 15 and 30) MRI scan magnetic field strength datasets. The 15 dataset contains 64 AD, 208 MCI and 90 NC; while the 30 dataset contains 66 AD and 110 NC. DARTEL VBM pipeline \cite{vbm_compute} is then implemented to preprocess the data. Finally, the input features consist of 2,527 8$\times$8$\times$8 mm$^{3}$ size voxels with average values in GM population template greater than 0.1. Experiments are designed on 15ADNC, 30ADNC and 15MCINC tasks.

\subsection{Prediction and Path Analysis}
\label{prediction}
10-fold cross-validation is adopted for classification evaluation. Under exactly the same experimental setup, comparison is made between GSplit LBI and other classifiers: SVM, MLDA (univariate model via t-test + LDA) \cite{Discriminativeanalysisof}, Graphnet \cite{graphnet}, Lasso \cite{lasso}, Elastic Net, TV+L$_{1}$ and $n^{2}$GFL. For each model, optimal parameters are determined by grid-search. For GSplit LBI, $\rho$ is chosen from $\{1,2,...,10\}$, $\kappa$ is set to 10; $\alpha = \nu / \kappa(1 + \nu \Lambda_{X}^{2} + \Lambda_{D}^{2})$\footnote{For logit model, $\alpha < \nu / \kappa(1 + \nu \Lambda_{H}^2 + \nu \Lambda_{X}^2)$ since $\Lambda_{X} > \Lambda_{H}$.}; specifically, $\nu$ is set to 0.2 (corresponding to $\nu \nrightarrow 0$ in Fig.~\ref{figure:2})\footnote{In this experiment, comparable prediction result will be given for $\nu \in (0.1,10)$.} . The regularization coefficient $\lambda$ is ranged in $\{0,0.05, 0.1,...,0.95,1,10,10^{2}\}$ for lasso\footnote{0 corresponds to logistic regression model.} and $2^{\{-20,-19,...,0,...,20\}}$ for SVM. For other models, parameters are optimized from $\lambda:\thinspace \{0.05,0.1,...,0.95,1,10,10^{2}\}$ and $\rho:  \thinspace \{0.5,1,..,10\}$(in addition, the mixture parameter $\alpha$: $\{0,0.05,...,0.95\}$ for Elastic Net).
\begin{table}
\caption{Comparison of GSplit LBI with other models}
\small
\centering

\begin{tabular}{c|c|c|c|c|c|c|c|c}
\hline
& MLDA & SVM & Lasso & Graphnet & Elastic Net & TV + $l_{1}$ & $n^{2}$GFL & GSplit LBI\thinspace($\beta_{pre}$) \\
\hline
15ADNC & $85.06\%$ & $83.12\%$ & $87.01\%$ & $86.36\%$ & $88.31\%$ & $83.77\%$ & $86.36\%$ & $\textbf{88.96\%}$ \\
\hline
30ADNC & $86.93\%$ & $87.50\%$ & $87.50\%$ & $88.64\%$ & $89.20\%$ & $87.50\%$ & $87.50\%$ & $\textbf{90.91\%}$  \\
\hline
15MCINC & $61.41\%$ & $70.13\%$ & $69.80\%$ & $72.15\%$ & $70.13\%$ & $73.83\%$ & $69.80\%$ & $\textbf{75.17\%}$ \\
\hline 
\end{tabular} 
\label{table:2}
\end{table}

The best accuracy in the path of GSplit LBI and counterpart are reported. Table~\ref{table:2} shows that $\beta_{pre}$ of our model outperforms that of others in all cases. 
Note that although our accuracies may not be superior to models with multi-modality data  \cite{Structuredsparsekernel}, they are the state-of-the-art results for only sMRI modality.

\begin{figure}
\includegraphics[width=1.0\columnwidth]{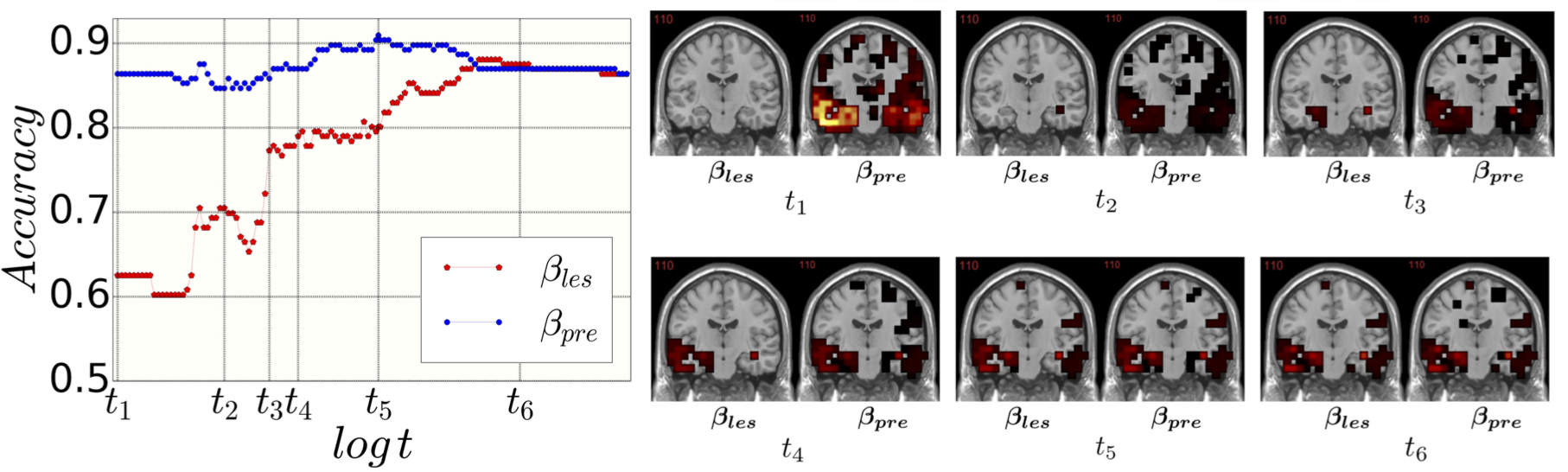}
\caption{Left image: Accuracy of $(\beta_{pre},\beta_{les})$ vs $log\thinspace t$\thinspace($t$: regularization parameter). Right image: Six 2-d brain slice images of selected degenerative voxels of $\beta_{les}$ and $\beta_{pre}$ are sorted orderly at $\{t_{1},...t_{6}\}$. As $t$ grows, $\beta_{pre}$ and $\beta_{les}$ identify similar lesion features.}\label{figure:5}
\end{figure}

The process of feature selection combined with prediction accuracy can be analyzed together along the path. The result of 30ADNC is used as an illustration in Fig.~\ref{figure:5}. We can see that $\beta_{pre}$ (blue curve) outperforms $\beta_{les}$ (red curve) in the whole path for additional procedural bias captured by $\beta_{pre}$. Specifically, at $\beta_{pre}$'s highest accuracy ($t_{5}$), there is a more than $8\%$ increase in prediction accuracy by $\beta_{pre}$. Early stopping regularization at $t_5$ is desired, as $\beta_{pre}$ converges to $\beta_{les}$ in prediction accuracy with overfitting when $t$ grows. Recall that positive (negative) features represent degenerate (enlarged) voxels. In each fold of $\beta_{pre}$ at $t_5$, the commonly selected voxels among top 150 negative (enlargement) voxels are identified as procedural bias shown in Fig.~\ref{figure:7}, where most of these GM voxels are enlarged and located near lateral ventricle or subarachnoid space etc., possibly due to enlargement of CSF space in those locations that are different from the lesion features.

\subsection{Lesion Features Analysis}
\label{lesion}
To quantitatively evaluate the stability of selected lesion features, multi-set Dice Coefficient\thinspace(mDC)\footnote{In \cite{n2gfl}, $mDC := \frac{10 | \cap_{k=1}^{10} S(k) | }{\sum_{k=1}^{10} \thinspace | S(k) |}$ where $S(k)$ denotes the support set of $\beta_{les}$ in k-th fold.} \cite{mdc,n2gfl} is applied as a measurement.  The 30ADNC task is again applied as an example, the mDC is computed for $\beta_{les}$ which achieves highest accuracy by 10-fold cross-validation. As shown from Table~\ref{table:4}, when $\nu = 0.0002$\thinspace(corresponding to $\nu \to 0$ in Fig.~\ref{figure:2}), the $\beta_{les}$ of our model can obtain more stable lesion feature selection results than other models with comparable prediction power. Besides, the average number of selected features (line 3 in Table~\ref{table:4}) are also recorded\thinspace. Note that although elastic net is of slightly higher accuracy than $\beta_{les}$, it selects much more features than necessary. 
\begin{table}
\caption{mDC comparison between GSplit LBI and other models}
\small
\begin{center}
\begin{tabular}{c|c|c|c|c|c|c}
\hline
 & Lasso & Elastic Net & Graphnet & TV + $l_{1}$ & $n^{2}$ GFL & GSplit LBI\thinspace($\beta_{les}$)\\
\hline
Accuracy & $87.50\%$ & $89.20\%$ & $88.64\%$ &  $87.50\%$ & $87.50\%$ &  $88.64\%$ \\
\hline
 mDC &0.1992 & 0.5631 & 0.6005 & 0.5824 & 0.5362 & \textbf{0.7805} \\
\hline
$\sum_{k=1}^{10} \thinspace | S(k) |/10$ & 50.2 & 777.8 & 832.6 & 712.6 & 443.9 & 129.4  \\
\hline
\end{tabular} 
\end{center}
\label{table:4}
\end{table}

\begin{figure}
\begin{center}
\begin{minipage}[t]{0.15\linewidth}
    \includegraphics[width= \columnwidth]{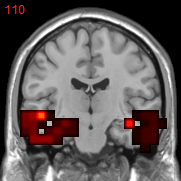}{\begin{center} (a) fold 2 \end{center}}
\end{minipage}
\begin{minipage}[t]{0.15\linewidth}
    \includegraphics[width= \columnwidth]{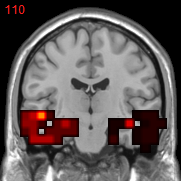}{\begin{center} (b) fold 10 \end{center}}
\end{minipage}
\begin{minipage}[t]{0.15\linewidth}
    \includegraphics[width= \columnwidth]{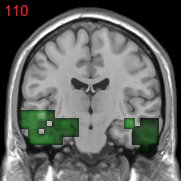}{\begin{center} (c) overlap \end{center}}
\end{minipage}
\begin{minipage}[t]{0.37\linewidth}
    \includegraphics[width= \columnwidth]{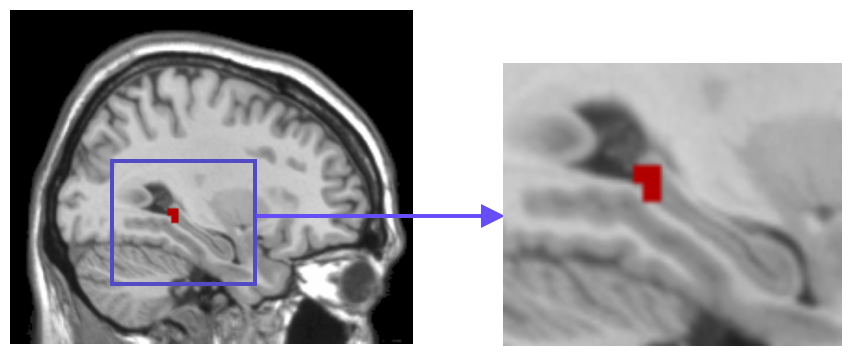}{\begin{center} (d) coarse-to-fine \end{center}}
\end{minipage}
\end{center}
\caption{(a)-(c): Stability of selected lesion features of $\beta_{les}$ shown in 2-d 110 slice brain images when $\nu = 0.0002$. (a)-(b): Results of fold 2 and fold 10. (c): The overlapped features in 10 folds. (d): The 2-d slice brain image of selected voxels with $2\times 2\times 2$ $mm^{3}$ using coarse-to-fine approach.}\label{figure:8}
\end{figure}

For the meaningfulness of selected lesion features, they are shown in Fig.~\ref{figure:8} (a)-(c), located in hippocampus, parahippocampal gyrus and medial temporal lobe etc., which are believed to be early damaged regions for AD patients. \newline

To further investigate the locus of lesion features, we conduct a coarse-to-fine experiment. Specifically, we project the selected overlapped voxels of $8 \times 8 \times 8$ $mm^{3}$ size (shown in Fig.~\ref{figure:8} (c)) onto MRI image with more finer scale voxels, i.e. in size of $2 \times 2 \times 2$ $mm^{3}$. Totally 4,895 voxels are served as input features after projection. Again, the GSplit LBI is implemented using 10-fold cross-validation. The prediction accuracy of $\beta_{pre}$ is $90.34\%$ and on average 446.6 voxels are selected by $\beta_{les}$. As desired, these voxels belong to parts of lesion regions, such as those located in hippocampal tail, as shown in Fig.~\ref{figure:8} (d).

\section{Conclusions}
In this paper, a novel iterative dual task algorithm is proposed to incorporate both disease prediction and lesion feature selection in neuroimage analysis. With variable splitting term, the estimators for prediction and selecting lesion features can be separately pursued and mutually monitored under a gap control. The gap here is dominated by procedural bias, some specific features crucial for prediction yet ignored in \emph{a priori} disease knowledge. With experimental studies conducted on 15ADNC, 30ADNC and 15MCINC tasks, we have shown that the leverage of procedural bias can lead to significant improvements in both prediction and model interpretability. In future works, we shall extend our model to other neuroimaging applications including multi-modality data.\newline

\setlength{\parindent}{0pt} \textbf{{Acknowledgements.}} This work was supported in part by 973-2015CB351800, 2015CB85600, 2012CB825501, NSFC-61625201, 61370004, 11421110001 and Scientific Research Common Program of Beijing Municipal Commission of Education (No. KM201610025013).

\bibliographystyle{splncs03}
\bibliography{paper_552_refer}

\begin{thebibliography}{10}
\providecommand{\url}[1]{\texttt{#1}}
\providecommand{\urlprefix}{URL }

\bibitem{vbm_compute}
Ashburner, J.: A fast diffeomorphic image registration algorithm. Neuroimage
  38(1),  95--113 (2007)

\bibitem{whyvoxel-based}
Ashburner, J., Friston, K.J.: Why voxel-based morphometry should be used.
  Neuroimage  14(6),  1238--1243 (2001)

\bibitem{Discriminativeanalysisof}
Dai, Z., Yan, C., Wang, Z., Wang, J., Xia, M., Li, K., He, Y.: Discriminative
  analysis of early alzheimer's disease using multi-modal imaging and
  multi-level characterization with multi-classifier. Neuroimage  59(3),
  2187--2195 (2012)

\bibitem{mdc}
Dice, L.R.: Measures of the amount of ecologic association between species.
  Ecology  26(3),  297--302 (1945)

\bibitem{graphnet}
Grosenick, L., Klingenberg, B., Katovich, K., Knutson, B., Taylor, J.E.:
  Interpretable whole-brain prediction analysis with graphnet. Neuroimage  72,
  304--321 (2013)

\bibitem{Splitlbi}
Huang, C., Sun, X., Xiong, J., Yao, Y.: Split lbi: An iterative regularization
  path with structural sparsity. advances in neural information processing
  systems. Advances In Neural Information Processing Systems pp. 3369--3377
  (2016)

\bibitem{bregman}
Osher, S., Ruan, F., Xiong, J., Yao, Y., Yin, W.: Sparse recovery via
  differential inclusions. Applied and Computational Harmonic Analysis  (2016)

\bibitem{Structuredsparsekernel}
Peng, J., An, L., Zhu, X., Jin, Y., Shen, D.: Structured sparse kernel learning
  for imaging genetics based alzheimer's disease diagnosis. International
  Conference on Medical Image Computing and Computer-Assisted Intervention pp.
  70--78 (2016)

\bibitem{TV}
Rudin, L.I., Osher, S., Fatemi, E.: Nonlinear total variation based noise
  removal algorithms. Physica D: Nonlinear Phenomena  60(1-4),  259--268 (1992)

\bibitem{lasso}
Tibshirani, R.: Regression shrinkage and selection via the lasso. Journal of
  the Royal Statistical Society. Series B (Methodological)  58,  267--288
  (1996)

\bibitem{Genlasso}
Tibshirani, R.J., Taylor, J.E., Candes, E.J., Hastie, T.: The solution path of
  the generalized lasso. The Annals of Statistic  39(3),  1335--1371 (2011)

\bibitem{vaiter}
Vaiter, S., Peyr{\'e}, G., Dossal, C., Fadili, J.: Robust sparse analysis
  regularization. IEEE Transactions on Information Theory  59(4),  2001--2016
  (2013)

\bibitem{n2gfl}
Xin, B., Hu, L., Wang, Y., Gao, W.: Stable feature selection from brain smri.
  AAAI pp. 1910--1916 (2014)

\bibitem{LBI}
Yin, W., Osher, S., Darbon, J., Goldfarb, D.: Bregman iterative algorithms for
  compressed sensing and related problems. SIAM Journal on Imaging Sciences
  1(1),  143--168 (2008)

\bibitem{elasticnet}
Zou, H., Hastie, T.: Regularization and variable selection via the elastic net.
  Journal of the Royal Statistical Society: Series B (Statistical Methodology)
  67(2),  301--320 (2005)

\end{thebibliography}

\appendix 
\newpage
\renewcommand\thesection{\Alph{section}}
\begin{center}
    \Large\bf{Supplementary Information}
\end{center}

\section{Notation}
For matrix $A$, $A_{J}$ represents the submatrix of $A$ indexed by $J$. $A^{\dag}$ denotes the Moore-Penrose pseudoinverse of $A$.  Suppose $A \in R^{n\times n}$, $\Vert A \Vert_{\Sigma} := \mathrm{trace}(A) =  \sum_{i=1}^{n} A_{i,i}$. Besides, $\tilde{\beta}$ and $\beta$ are used to represent $\beta_{les}$ and $\beta_{pre}$ respectively in what follows.

\section{Model selection consistency}

Consider recovery from generalized linear model(GLM) of $\beta^{\star} \in R^{p}$ which satisfies structural sparsity after linearly transformed by $D \in R^{m \times p}$:
\begin{eqnarray}
& P(y | x,\beta^{\star})  \propto   exp(\frac{x^T\beta^{\star} \cdot y - \psi(x^T\beta^{\star})}{d(\sigma)} ) \nonumber \\
 s.t. \quad & \gamma^{\star} = D\beta^{\star}\thinspace   (S := supp(\gamma^{\star}),  \thinspace s = |S|,\thinspace s<<m)
 \label{eq:1}
\end{eqnarray}
where $\psi: R \to R$ is link function and $d(\sigma)$ is known parameter related to the variance of distribution. 

Under linear model with $\psi(t) = t^{2}$ and $d(\sigma) = \sigma^{2}$ in~\ref{eq:1}, our model GSplit LBI degenerates to Split LBI \cite{Splitlbi}. Recently, it's proved in \cite{Splitlbi} that the Split LBI may achieve model selection consistency under weaker conditions than generalized lasso \cite{Genlasso,vaiter} if $\nu$ is large enough. We claim that this property can also be shared by logit model. 

To understand why Gsplit LBI can achieve better model selection consistency, note that the variable splitting term projects solution vector $\beta$ into higher dimensional space $(\beta,\gamma)$ with $\beta$ fitting data and $\gamma$ being structural sparse. This will make it easier for the subspace of $\gamma_{S^{c}}$ to decorrelate with the subspace of $(\beta,\gamma_{S})$, especially when $\nu$ increases, which sheds light on better performance of Split LBI to recover true signal set $S$. What's more important, the property may also be shared by logit model when $y = \{\pm 1\}$, $d(\sigma) = 1$ and $\psi(t) = log(1 + exp(t))$. Concretely speaking, we use $\theta_{S^{c},(\beta,S)}(\nu)$ to denote the angle between subspace of $\gamma_{S^{c}}$ and that of $(\beta,\gamma_{S})$, the definition of which is:
\begin{equation}
\theta_{S^{c},(\beta,S)}(\nu) := \mathrm{arccos}\thinspace (\frac{\Vert P_{A_{(\beta,S)}}A_{S^{c}} \Vert_{F}}{\Vert A_{S^{c}} \Vert_{F}}) =  \mathrm{arccos}\thinspace (\sqrt{\frac{\Vert H_{S^{c},(\beta,S)}H^{\dagger}_{(\beta,S),(\beta,S)}H_{(\beta,S),S^{c}} \Vert_{\Sigma}}{\Vert H_{S^{c},S^{c}} \Vert_{\Sigma}}})
\label{eq:2}
\end{equation}
Where $A := \left[ \begin{array}{cc} A_{(\beta,S)} & A_{S^{c}} \end{array} \right]$ and $H := \triangledown^{2}_{\beta,\gamma} \thinspace l(\beta,\gamma) = A^TA = \left[ \begin{array}{cc} H_{(\beta,S),(\beta,S)} & H_{(\beta,S),S^{c}} \\ H_{S^{c},(\beta,S)} & H_{S^{c},S^{c}} \end{array} \right]$. \newline
\begin{remark}
For linear model, $A = \triangledown_{\beta,\gamma}\thinspace l(\beta,\gamma)$ with 
\begin{eqnarray}
A_{(\beta,S)} = \left[ \begin{array}{cc} X & 0_{n \times s} \\  -D_{S} / \sqrt{\nu} & I_{(S,S)} / \sqrt{\nu} \\ -D_{S^{c}} / \sqrt{\nu} & 0_{(p-s) \times s} \end{array} \right] \quad A_{S^{c}} = \left[\begin{array}{cc} 0_{n \times (p-s)} \\ 0_{s \times (p-s)} \\ I_{(S^{c},S^{c})} / \sqrt{\nu} \end{array} \right] 
\label{eq:3}
\end{eqnarray}
There is no explicit definition for $A$ for logit model, however $\theta_{S^{c},(\beta,S)}(\nu)$ can be computed through Hessian matrix $H$ in equation~\ref{eq:3}.
\end{remark}
We claim that $\theta_{S^{c},(\beta,S)}(\nu)$ will increase as $\nu$ becomes larger under some conditions. See theorem~\ref{theo:1} for details.
\begin{theorem}
\label{theo:1}
Under linear model and logit model, $lim_{\nu \to +\infty} \theta_{S^{c},(\beta,S)}(\nu) = 90^{\circ}$ if and only if $\mathrm{Im}(D_{S^{c}}^T) \subseteq \mathrm{Im}(X^T)$.
\end{theorem}
\begin{remark}
In \cite{Splitlbi}, it's been proved that the necessary condition for sign-consistency is $IRR(\nu) < 1$. For uniqueness of model, we also assume that $ker(X)$ $\cap$ $ker(D_{S^{c}})$ $\subseteq ker(D_{S})$. Combined with $\mathrm{Im}(D_{S^{c}}^T) \subseteq \mathrm{Im}(X^T) \iff ker(X) \subseteq ker(D_{S^c})$, we have that $ker(X) \subseteq ker(D_{S})$, which is the sufficient and necessary condition for the hold of $\lim_{\nu \to \infty} \mathrm{IRR}(\nu) \to 0$. Hence, this is another way to understand why GSplit LBI can achieve better model selection consistency.
\end{remark}
\begin{proof}
We firstly prove the case under linear model. Denoted $A: = \nu X^{\star}X + D^TD$ where $X\in R^{n \times p}$ and $X^{\star} = X/n$. Note that:
\begin{eqnarray*}
H_{(\beta,S),(\beta,S)} = QLQ^T, \quad H_{S^{c},(\beta,S)} = \left[ \begin{array}{cc} D_{S^{c}}/\nu & 0 \end{array} \right]
\end{eqnarray*}
where:
\begin{eqnarray}
Q = \left[ \begin{array}{cc} I_{p} & 0 \\ -D_{S}A^{\dagger} & I_{s} \end{array} \right], \quad L = \left[ \begin{array}{cc} A/\nu & 0 \\ 0 & (I_{s} - D_{S}A^{\dagger}D_{S}^T)/ \nu \end{array}\right]
\end{eqnarray}
Then we have:
\begin{equation}
H_{S^{c},(\beta,S)}H^{\dagger}_{(\beta,S),(\beta,S)}H_{(\beta,S),S^{c}} = H_{S^{c},(\beta,S)}QL^{\dagger}Q^TH_{(\beta,S),S^{c}} = \frac{1}{\nu} D_{S^{c}}A^{\dagger}D_{S^{c}}^T
\label{eq:4}
\end{equation}
Substituting equation~\ref{eq:4} into the second equation of~\ref{eq:2}, we have:
\begin{equation}
\mathrm{cos}^{2}(\theta_{S^{c},(\beta,S)}(\nu)) = \frac{\Vert D_{S^{c}}A^{\dagger}D_{S^{c}} \Vert_{\Sigma}}{\Vert H_{S^{c},S^{c}} \Vert_{\Sigma}}  = \frac{\Vert D_{S^{c}}A^{\dagger}D_{S^{c}} \Vert_{\Sigma}}{m - s} 
\label{eq:5}
\end{equation}
Denote $e_{i} \in R^{m-s}$ as the vector with the $i^{th}$ element being 1 and left being 0. Then equation~\ref{eq:5} is equivalent to:
\begin{equation}
\mathrm{cos}^{2}(\theta_{S^{c},(\beta,S)}(\nu))(m-s) = \Sigma_{i=1}^{p} d_{i}^TA^{\dagger}d_{i} 
\label{eq:6}
\end{equation}
where $d_{i} := D_{S^{c}}^Te_{i}$. Suppose the compact singular value decomposition of $X/\sqrt{n} := U\Lambda V^T$, and $(V, \tilde{V})$ be an orthogonal square matrix. Suppose the compact singular value decomposition of $D\tilde{V} := U_{1}\Lambda_{1}V_{1}^T$. If $\mathrm{Im}(D_{S^{c}}^T) \subseteq \mathrm{Im}(X^T) $, then $\exists f_{i}$, such that $d_{i} = Vf_{i}$, hence, 
\begin{align}
d_{i}^T (\nu X^{\star}X + D^TD)^{\dagger} d_{i} & = d_{i}^T\begin{pmatrix} V & \tilde{V} \end{pmatrix} \left( \begin{pmatrix} V^T \\ \tilde{V}^T \end{pmatrix} (\nu X^{\star}X + D^TD) \begin{pmatrix} V & \tilde{V} \end{pmatrix} \right)^{\dag} \begin{pmatrix} V^T \\ \tilde{V}^T \end{pmatrix} d_{i}  \nonumber \\
\label{eq:7}
& = f_{i}^T (\nu \Lambda^2 + V^TD^TDV)^{-1} f_{i} \to 0, \ as \ \nu \to \infty
\end{align}
Combined with equation~\ref{eq:6}, it's then easy to obtain that $\mathrm{cos}^{2}(\theta_{S^{c},(\beta,S)}(\nu)) \to 0$ as $\nu \to +\infty$. On the contrary, if $\exists \thinspace a$ such that $D_{S^{c}}^Ta \notin \mathrm{Im}(X^T)$, then there $\exists \thinspace i^{\star}$ such that $d_{i^{\star}} \notin \mathrm{Im}(X^T)$. This means that for $d_{i^{\star}}$, there $\exists \thinspace f_{1,i^{\star}}, f_{2,i^{\star}} \neq 0$ such that $d_{i^{\star}} = Vf_{1,i^{\star}} + \tilde{V}f_{2,i^{\star}}$. Then we have 
\begin{equation*}
d_{i^{\star}}^T (\nu X^{\star}X + D^TD)^{\dagger} d_{i^{\star}}  \geq f_{2,i^{\star}}^T (\tilde{V}^TD^TD\tilde{V})^{\dag}f_{2,i^{\star}} = f_{2,i^{\star}}^TV_{1}\Lambda_{1}^{-2}V_{1}^Tf_{2,i^{\star}}
\end{equation*}
does not equal to $0 \iff f_{2,i^{\star}}^TV_{1}\Lambda_{1}^{2}V_{1}^Tf_{2,i^{\star}} = f_{2,i^{\star}}^T \tilde{V}^TD^TD\tilde{V} f_{2,i^{\star}} \neq 0$. Since
\begin{equation*}
f_{2,i^{\star}}^T \tilde{V}^TD^TD\tilde{V} f_{2,i^{\star}} \geq f_{2,i^{\star}}^T \tilde{V}^T d_{i^{\star}}d_{i^{\star}}^T f_{2,i^{\star}} = (f_{2,i^{\star}}^Tf_{2,i^{\star}})^2 > 0
\end{equation*}
From equation~\ref{eq:6}, we can obtain that:
\begin{equation*}
\mathrm{cos}^{2}(\theta_{S^{c},(\beta,S)}(\nu))(m-s) \geq d_{i^{\star}}^TA^{\dagger}d_{i^{\star}}  \neq 0
\end{equation*}
which means the $\theta_{S^{c},(\beta,S)}(\nu) \to 0$ does not hold when $\nu \to +\infty$. The proof is then completed under linear model. Under logit model, the definition of $A$ is modified to $A: = \nu X^{\star}W(\{x_{i},\beta\}_{i=1}^{p})X + D^TD$ where $W(\{x_{i},\beta\}_{i=1}^{p})$ is a diagonal matrix with each diagonal element equals to $\frac{exp(x_{i}^T\beta)}{(1 + exp(x_{i}^T\beta))^{2}}$, the left proof is almost the same with that of linear model.
\end{proof}

\begin{figure}[!h]
\begin{center}
\begin{minipage}[t]{0.59\linewidth}
    \includegraphics[width= \columnwidth]{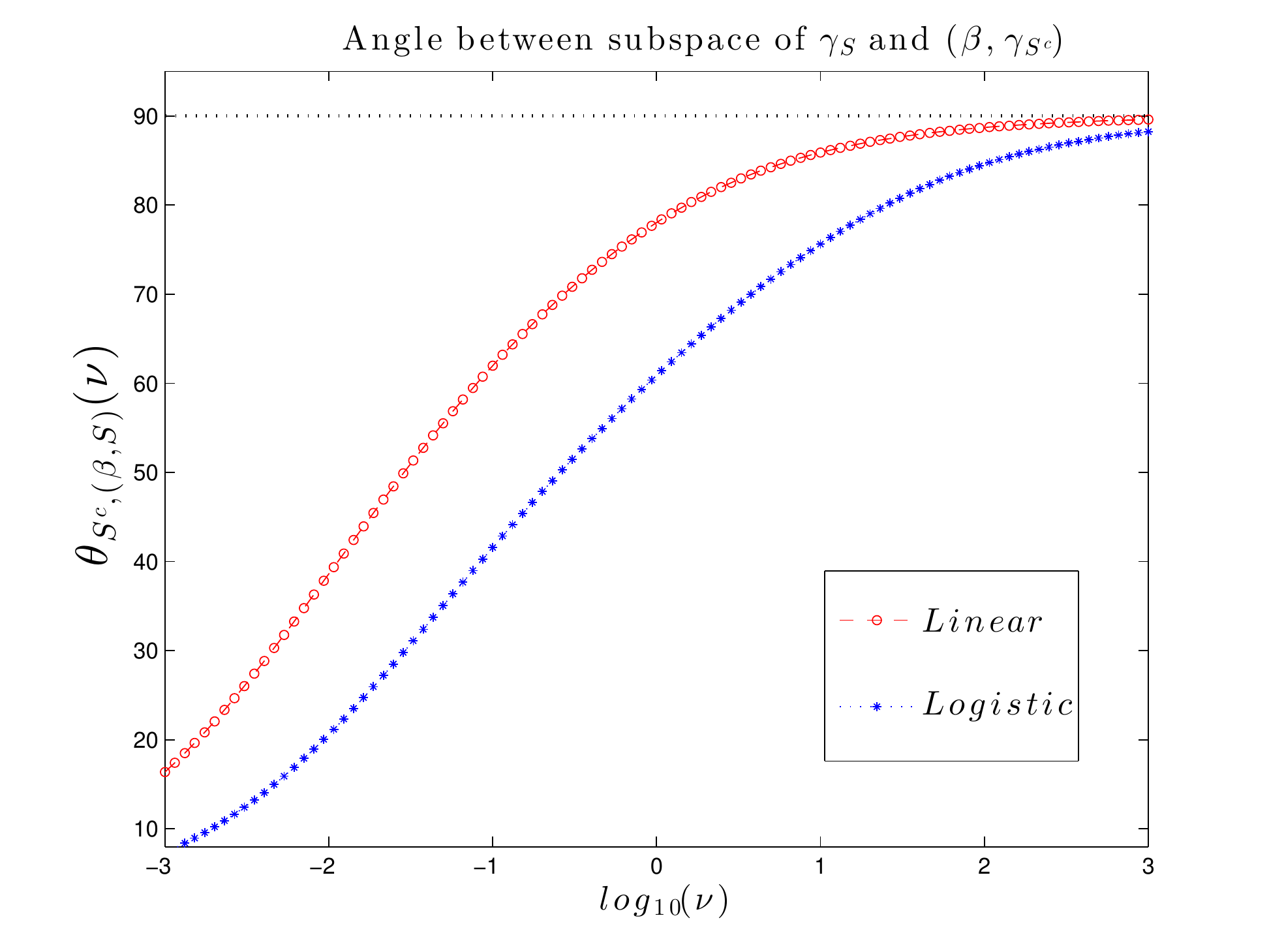}
\end{minipage}
\begin{minipage}[t]{0.35\linewidth}
    \includegraphics[width= 1.1\columnwidth]{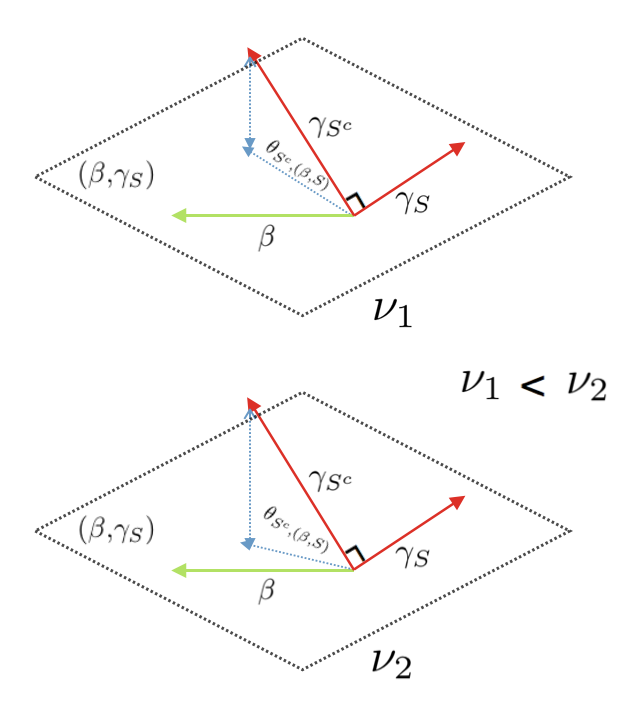}
\end{minipage}	
\end{center}
\caption{Left image: The $\theta_{S^{c},(\beta,S)}(\nu)$ curve of logit model and linear model. Right image: Illustration of $\theta_{S^{c},(\beta,S)}(\nu)$ and it monotonically increase v.s. $\nu$.}\label{figure:1}
\end{figure}

 An simulation experiment is conducted to illustrate this idea. In more detail, $n = 100$ and $p = 80$, $D = I$ and $X \in R^{n \times p}$ and $X_{i,j} \sim N(0,1)$. $\beta^{\star}_{i} =  2$ for $1 \leq i \leq 4$, $\beta^{\star}_{i} =  -2$ for $5 \leq i \leq 8$ and 0 otherwise, $y$ is generated by both linear model $y = X\beta^{\star} + \epsilon$ with $\epsilon \sim N(0,1)$ and logit model given $X$ and $\beta^{\star}$. We simulated for 100 times and average $\theta_{S^{c},(\beta,S)}(\nu)$ is then computed, which is shown in the left image in figure~\ref{figure:1}. We can see that $\theta_{S^{c},(\beta,S)}(\nu)$ increases when $\nu$ becomes larger, as illustrated in right image in figure~\ref{figure:1}, and converges to $90^\circ$ when $\nu \to +\infty$.

The average AUC and estimation of $\beta^{\star}$ of Gsplit LBI with different $\nu$ compared with those of genlasso are also computed. Table~\ref{table:1} shows better AUC with the increase of $\nu$ before $\nu = 100$. As we can see from the algorithm in the paper that $\tilde{\beta}$ is the projection of $\beta$ onto the support set of $\gamma$. Hence it is equivalent to say that better model selection of $\tilde{\beta}$ can be achieved as $\nu$ increases.

However, the excessively large value of $\nu$ will lower the signal-to-noise ratio, which is also crucial for model selection consistency and prediction estimation. It's shown in \cite{Splitlbi} that $\nu$ determines the trade-off between model selection consistency and estimation of $\beta^{\star}$. Also, the irrepresentable condition(IRR) can be satisfied as long as $\nu$ is large enough. If $\nu$ continuously increase, it will deteriorate the estimation of $\beta^{\star}$, prediction estimation and even AUC. In our experiment the same phenomena can be observed, i.e. the estimation of $\tilde{\beta}$ and $\beta$ get worse if $\nu$ increases from 10 and 100, respectively; when $\nu = 100$, AUC even decreases. 
\begin{table}[!h]
\begin{center}
\caption{Comparison between Gsplit LBI with different $\nu$ and genlasso in terms of AUC, $\Vert \tilde{\beta} - \beta^{\star} \Vert_{2}$, $\Vert \beta - \beta^{\star} \Vert_{2}$, $\Vert X\tilde{\beta} - X\beta^{\star} \Vert_{2}$ and $\Vert X\beta - X\beta^{\star} \Vert_{2}$.}
\small
\begin{tabular}{c|c|c|c|c|c|c|c}
\hline
Model & \multicolumn{6}{|c|}{Gsplit LBI} & genlasso \\
\hline 
$\nu$ & 0.02 & 0.1 & 1 & 5 & 10 & 100 & - \\
\hline
AUC & 0.9531 & 0.98194 & 0.98514 & 0.98791 & 0.98792 & 0.98590 & 0.97915 \\
\hline
$\Vert \tilde{\beta} - \beta^{\star} \Vert_{2}$ & 4.9079 & 4.9015 & 4.8513 & 4.8495 & 4.8473 & 5.3578 & - \\
\hline
$\Vert \beta - \beta^{\star} \Vert_{2}$ & 3.9187 & 3.8993 & 3.7619 & 3.6814 & 3.7129 & 5.1540 & 4.9113 \\
\hline
$\Vert X\tilde{\beta} - X\beta^{\star} \Vert_{2}$ & 47.0224 & 46.9625 & 46.6784 & 46.9158 & 47.0098 & 52.2737 & - \\
\hline
$\Vert X\beta - X\beta^{\star} \Vert_{2}$ & 37.3496 & 37.0962 & 35.4987 & 34.6423 & 34.8514 & 49.61020 & 50.6408 \\
\hline
\end{tabular} 
\end{center}
\label{table:1}
\end{table}

\section{Relationship between $\beta$ and $\tilde{\beta}$}
The estimate $\tilde{\beta}$, as a projection of $\beta$ onto the subspace of $\gamma$, can select features that satisfy structural sparsity. Following the Linearized Bregman Iteration \cite{LBI}, $\beta$ and $\tilde{\beta}$ will be more similar on features selected by $\tilde{\beta}$. In more detail, note that when $t = 0$, $\tilde{\beta}(t) = 0$ and $\beta(t)$ is the graph laplacian regularizer with penalty factor $\frac{1}{2\nu}$. As $t$ progresses, the gap between $\beta(t)$ and $\tilde{\beta}(t)$ will decrease in terms of $\Vert \beta(t) - \tilde{\beta}(t) \Vert_{2}$ for every $\nu$, as shown in figure~\ref{figure:2}.
\begin{figure}[!h]
\centering
\begin{minipage}{0.32\linewidth}
	\includegraphics[width= \columnwidth]{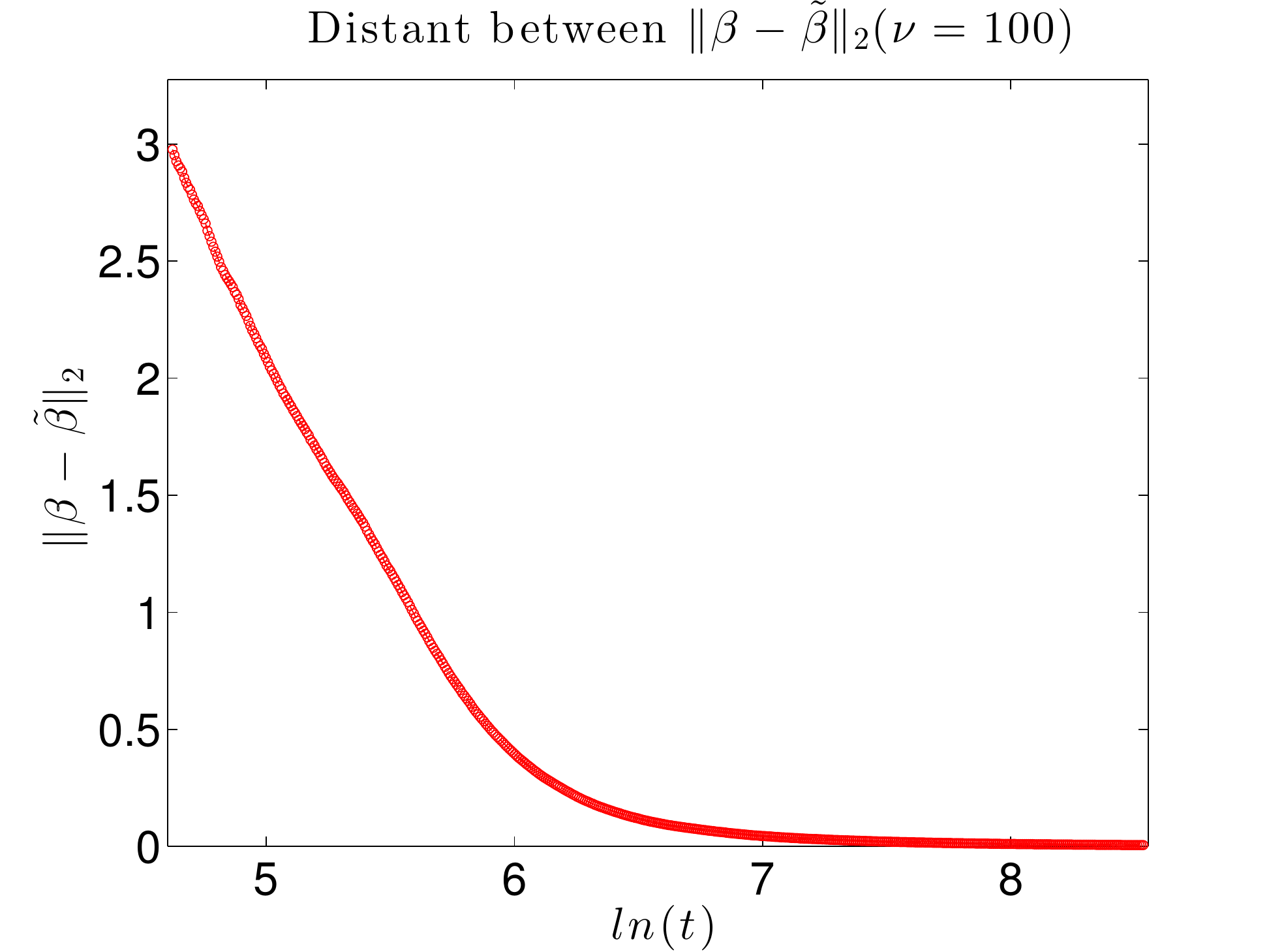}
\end{minipage}
\begin{minipage}{0.32\linewidth}
	\includegraphics[width= \columnwidth]{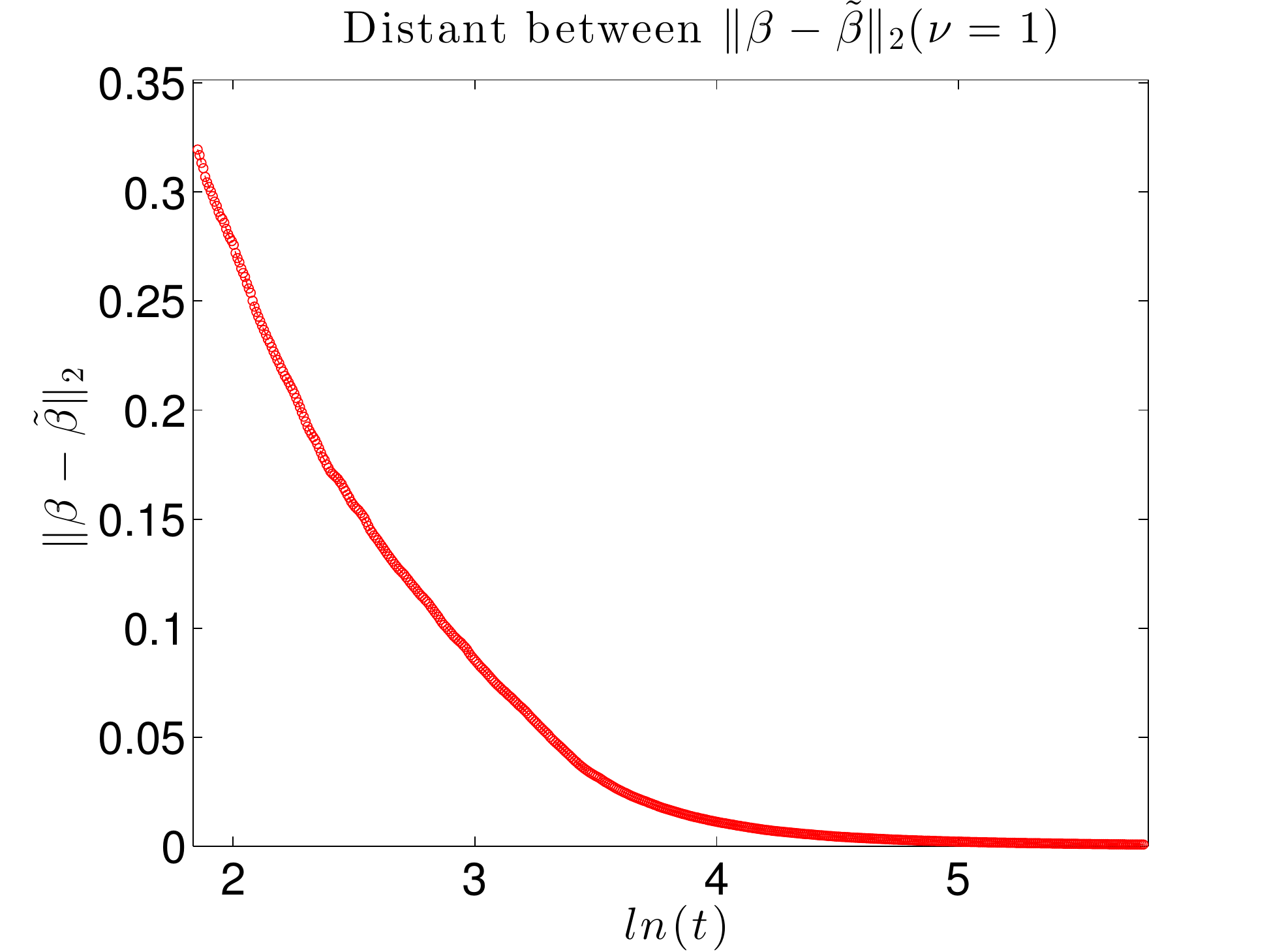}
\end{minipage}
\begin{minipage}{0.32\linewidth}
	\includegraphics[width= \columnwidth]{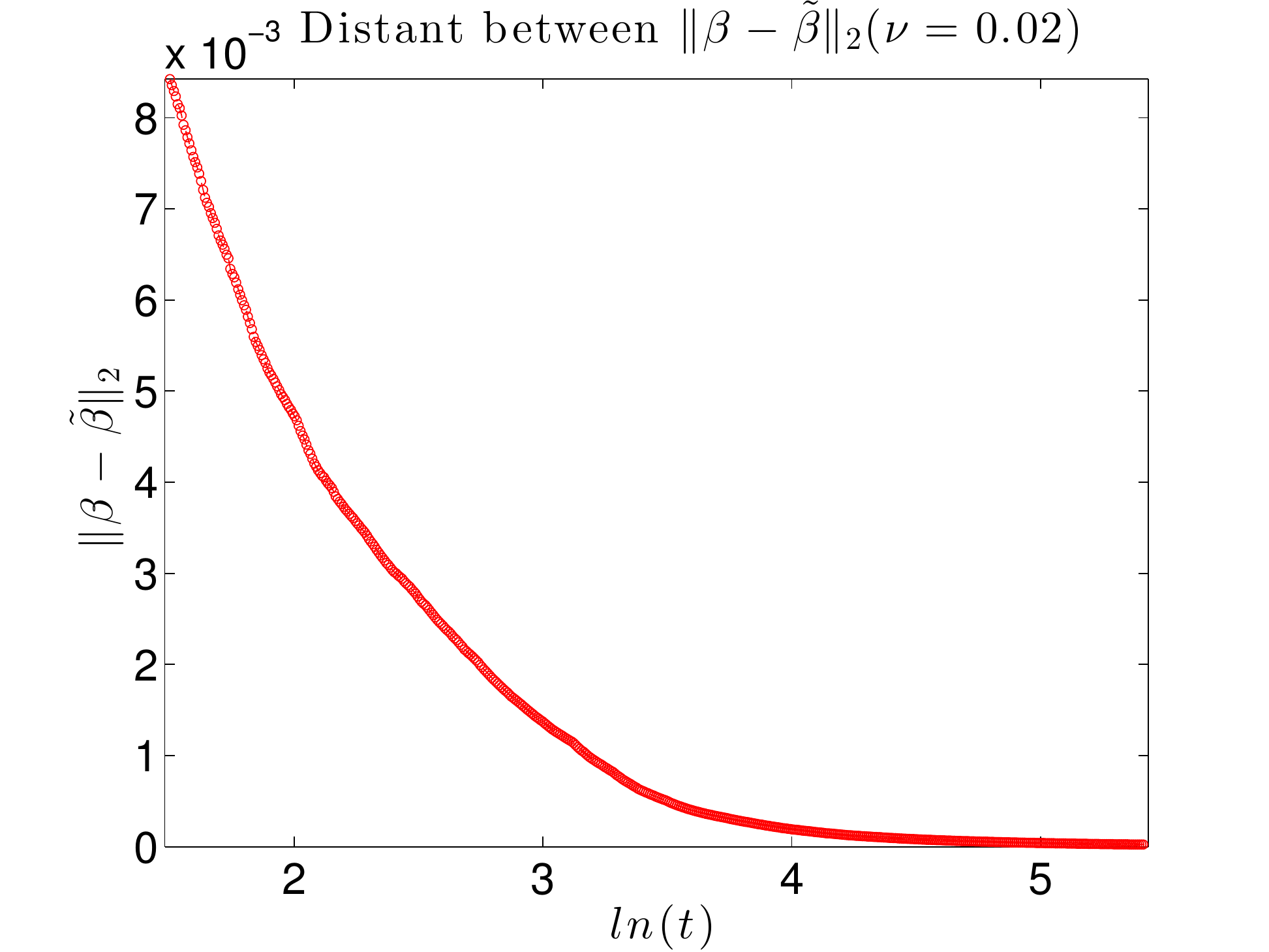}
\end{minipage}
\caption{$\Vert \beta - \tilde{\beta} \Vert_{2}$ in the regularized solution path when $\nu = 100,1,0.02$. As $\nu$ decreases, the distance of $\beta(t)$ and $\tilde{\beta}(t)$ are tended to be with smaller distance.}
\label{figure:2}
\end{figure}

\begin{figure}[!h]
\centering
\begin{minipage}{0.32\linewidth}
	\includegraphics[width= \columnwidth]{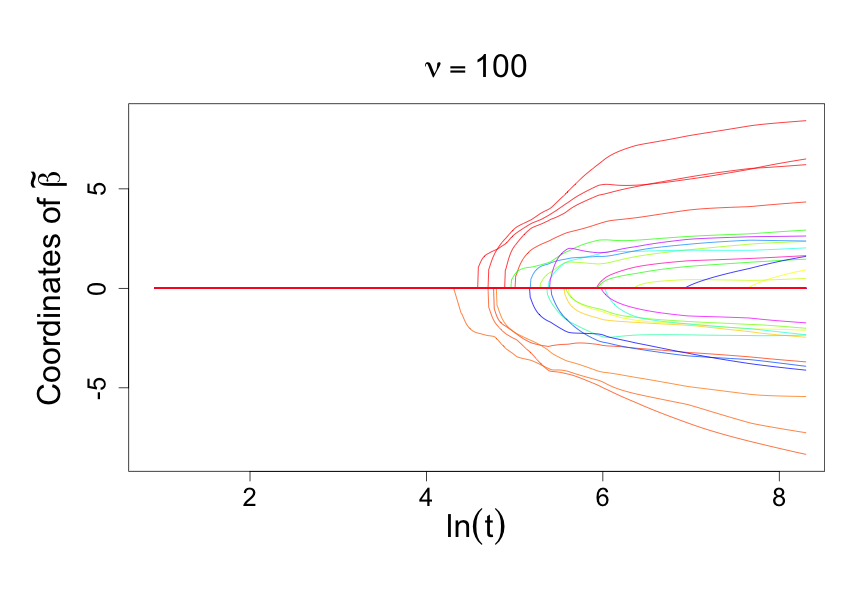}
\end{minipage}
\begin{minipage}{0.32\linewidth}
	\includegraphics[width= \columnwidth]{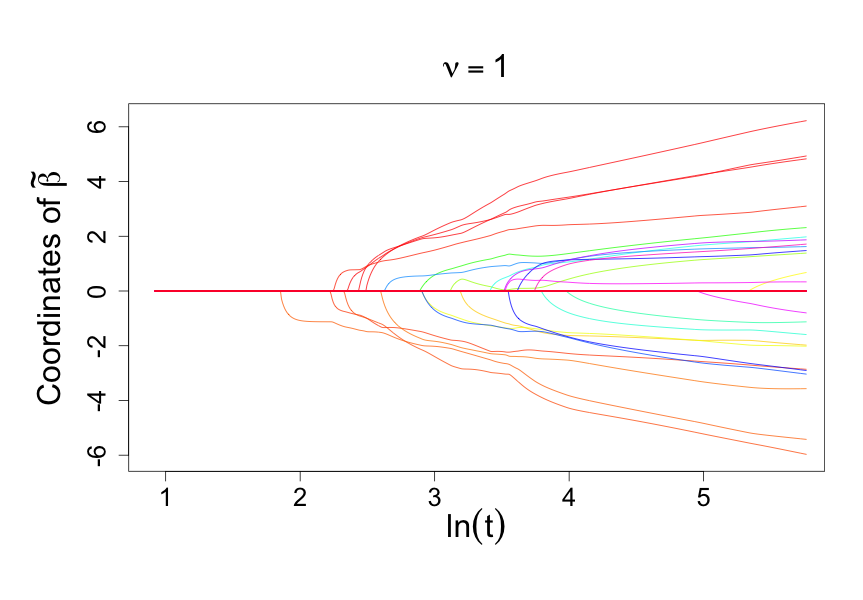}
\end{minipage}
\begin{minipage}{0.32\linewidth}
	\includegraphics[width= \columnwidth]{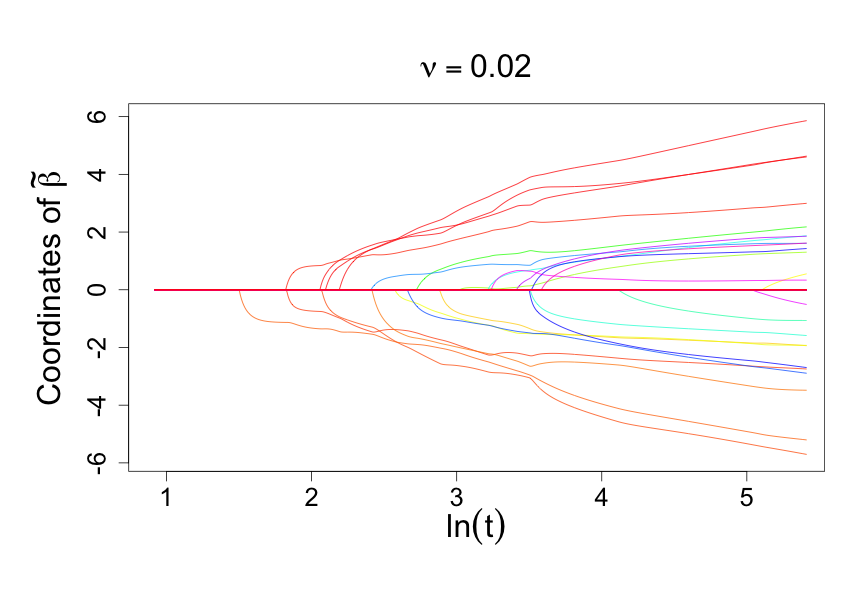}
\end{minipage}
\begin{minipage}{0.32\linewidth}
	\includegraphics[width= \columnwidth]{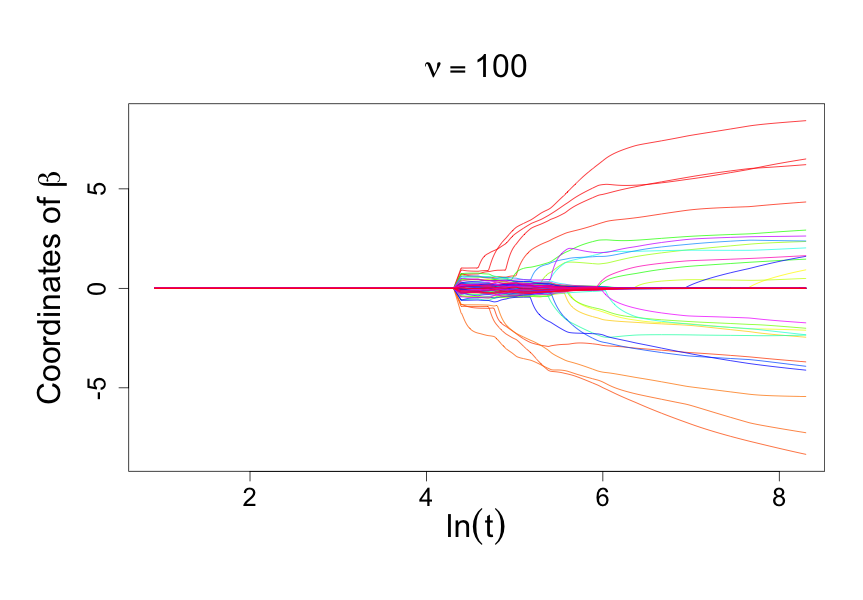}
\end{minipage}
\begin{minipage}{0.32\linewidth}
	\includegraphics[width= \columnwidth]{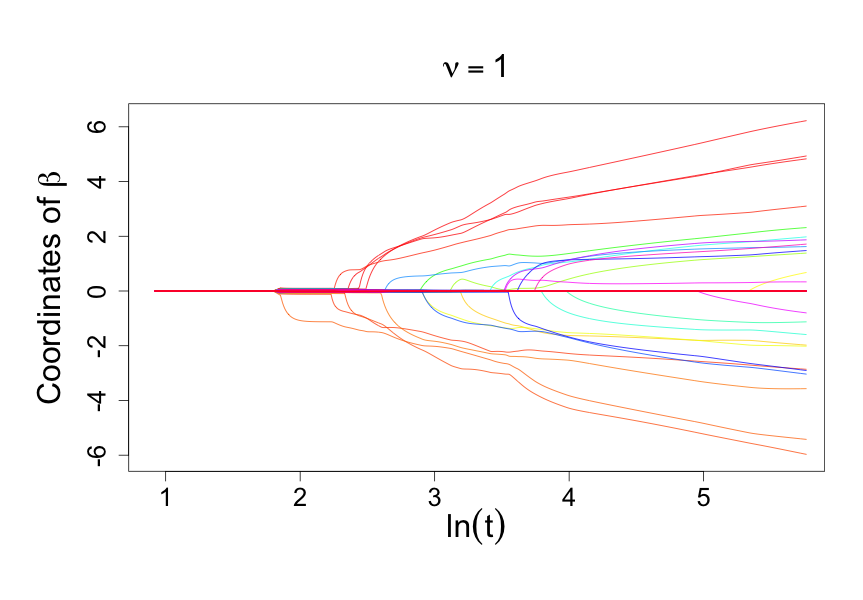}
\end{minipage}
\begin{minipage}{0.32\linewidth}
	\includegraphics[width= \columnwidth]{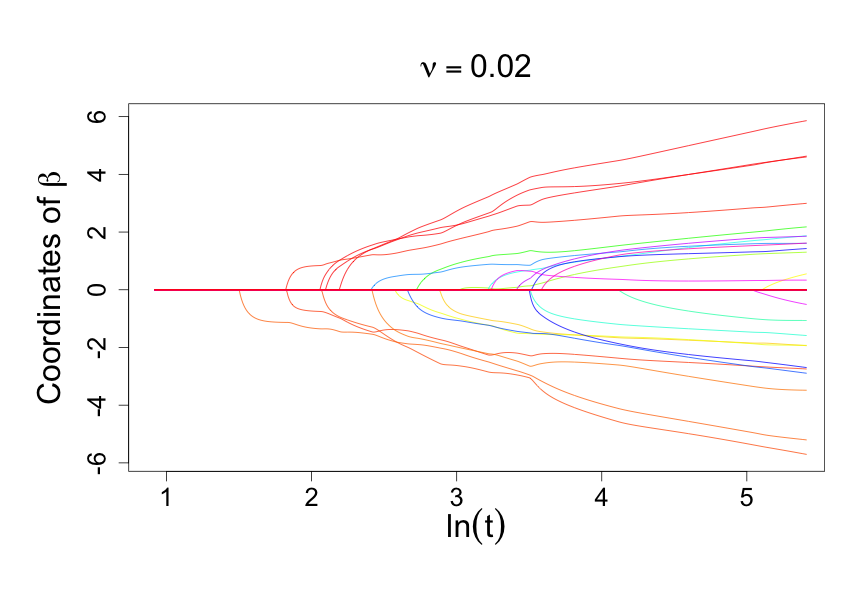}
\end{minipage}
\caption{Comparison of regularized solution path between $\beta$ and $\tilde{\beta}$ when $\nu = 100,1,0.02$. They look more similar with each other as $\nu$ decreases.}
\label{figure:3}
\end{figure}
 
Since $\Vert \beta(t) - \tilde{\beta}(t) \Vert_{2} \to 0$ as $t \to +\infty$ and $\tilde{\beta}$ is sparse, it follows that $\beta$ will approximate to $\tilde{\beta}$ on those selected features. In addition to these selected features, before convergence to $\tilde{\beta}$, $\beta$ can capture other features to better fit data(minimize training error), especially for those ones that significantly correlated with data. 

\section{Choice of $\nu$}

The choice of $\nu$ is task-dependent. For stable feature selection,  $\nu$ with rather "small" value is suggested. It's noted that $\beta - \tilde{\beta} \to 0$ as $\nu \to 0^{+}$, which is reflected by $l_{2}$ norm and regularized solution path shown in figure~\ref{figure:2}, \ref{figure:3}. In this case, the estimator $\tilde{\beta}$ will be constrained in comparably lower dimension space, therefore it may fit data with more stability, notwithstanding $\beta$ have no ability to select other features. 

For prediction estimation, the appropriately large value of $\nu$ is preferred. On one hand, when $\nu$ is appropriately large, the ability of selecting features with better model selection consistency can be achieved and $\beta$ will share closer values on these selected features as $t$ progress, as shown in figure~\ref{figure:3}. On the other hand, $\beta$ may increase the ability of fitting data by having other features being non-zeros as long as $\nu$ is not too small. In fact, it is shown in table~\ref{table:1} that comparable results can be given as long as $\nu$ belongs to a reasonable range of values(0.1-10 in this case).  

\section{IDS of ADNI subject used in our experiments}
\small
\begin{center}
\begin{longtable}{c|c|c||c|c|c||c|c|c}
 \toprule
 Subject & ID & Class &  Subject & ID & Class &  Subject & ID & Class  \\
 \midrule
123\textunderscore S\textunderscore0094	 & 	9655	 & 	15AD   &  027\textunderscore S\textunderscore   0408 &      14964         &       15MCI   & 072\textunderscore S\textunderscore0315  &     12559          &       15NC \\

123\textunderscore S\textunderscore 0088	 & 	9788	 & 	15AD   &  137\textunderscore S\textunderscore    0481 &      15044         &       15MCI & 137\textunderscore S\textunderscore   0301 &       12584        &       15NC \\

098\textunderscore S\textunderscore 0149	 & 	10146	 & 	15AD  &  027\textunderscore S\textunderscore   0417  &    15148           &       15MCI & 002\textunderscore S\textunderscore  0295  &     13722          &       15NC \\

032\textunderscore S\textunderscore 0147	 & 	10404	 & 	15AD  &  053\textunderscore S\textunderscore  0507  &      15315         &       15MCI	 &  037\textunderscore S\textunderscore 0327   &   13802            &       15NC \\

123\textunderscore S\textunderscore 0162	 & 	10962	 & 	15AD   & 094\textunderscore S\textunderscore   0531 &       15431        &       15MCI &  027\textunderscore S\textunderscore  0403  &   14146            &       15NC \\

128\textunderscore S\textunderscore 0216	 & 	11101	 & 	15AD  & 033\textunderscore S\textunderscore   0567 &       15459        &       15MCI & 137\textunderscore S\textunderscore  0459  &    14178           &       15NC \\

128\textunderscore S\textunderscore 0167	 & 	11203	 & 	15AD & 127\textunderscore S\textunderscore   0394  &        15510       &       15MCI & 002\textunderscore S\textunderscore   0413 &     14437          &       15NC \\

005\textunderscore S\textunderscore 0221	 & 	11604	 & 	15AD & 033\textunderscore S\textunderscore  0514  &        15605       &       15MCI	& 068\textunderscore S\textunderscore 0473    &     14483          &       15NC \\

014\textunderscore S\textunderscore 0328	 & 	12327	 & 	15AD & 033\textunderscore S\textunderscore    0513 &         15622      &       15MCI	& 116\textunderscore S\textunderscore 0360   &    14623           &       15NC \\

007\textunderscore S\textunderscore 0316	 & 	12616	 & 	15AD & 130\textunderscore S\textunderscore   0460  &        15711       &       15MCI & 133\textunderscore S\textunderscore  0488  &    14838           &       15NC \\

021\textunderscore S\textunderscore 0343	 & 	12979	 & 	15AD & 098\textunderscore S\textunderscore   0542 &         15848      &       15MCI	&133\textunderscore S\textunderscore  0493  &     14848          &       15NC \\

014\textunderscore S\textunderscore 0356	 & 	13004	 & 	15AD & 007\textunderscore S\textunderscore   0414 &         15875     &       15MCI	&014\textunderscore S\textunderscore 0520   &    15299           &       15NC \\

032\textunderscore S\textunderscore 0400	 & 	13525	 & 	15AD & 031\textunderscore S\textunderscore  0568  &         15885       &       15MCI	& 014 \textunderscore S\textunderscore  0519  &     15323          &       15NC \\

116\textunderscore S\textunderscore 0370	 & 	14122	 & 	15AD & 037\textunderscore S\textunderscore  0501  &         15916      &       15MCI & 116 \textunderscore S\textunderscore  0382  &      15347         &       15NC \\

127\textunderscore S\textunderscore 0431	 & 	15497	 & 	15AD & 037\textunderscore S\textunderscore  0552  &        15970       &       15MCI & 128\textunderscore S\textunderscore  0500  &  15366             &       15NC \\

031\textunderscore S\textunderscore 0554	 & 	15994	 & 	15AD & 130\textunderscore S\textunderscore   0423 &         16196      &       15MCI &  010\textunderscore S\textunderscore  0419  &    15415           &       15NC \\

128\textunderscore S\textunderscore 0517	 & 	16150	 & 	15AD & 014\textunderscore S\textunderscore  0557  &        16304       &       15MCI &  131\textunderscore S\textunderscore  0436  &    15674           &       15NC \\

116\textunderscore S\textunderscore 0487	 & 	16377	 & 	15AD & 033\textunderscore S\textunderscore 0511   &      16314         &       15MCI &128\textunderscore S\textunderscore 0522   & 15821              &       15NC \\

002\textunderscore S\textunderscore 0619	 & 	16392	 & 	15AD & 130\textunderscore S\textunderscore 0449   &      16351         &       15MCI & 033\textunderscore S\textunderscore  0516  &    15860           &       15NC \\

131\textunderscore S\textunderscore 0497	 & 	16666	 & 	15AD & 027\textunderscore S\textunderscore 0461   &      16467         &       15MCI	&  002\textunderscore S\textunderscore  0559  &    15948           &       15NC \\

021\textunderscore S\textunderscore 0642	 & 	17632	 & 	15AD & 128\textunderscore S\textunderscore 0608   &     16503          &       15MCI	& 014\textunderscore S\textunderscore  0548  &  16024             &       15NC  \\

033\textunderscore S\textunderscore 0739	 & 	19175	 & 	15AD & 128\textunderscore S\textunderscore  0611  &      16766         &       15MCI  & 128\textunderscore S\textunderscore  0545  &     16090          &       15NC \\

100\textunderscore S\textunderscore 0743	 & 	19585	 & 	15AD & 053\textunderscore S\textunderscore  0621  &      16864         &       15MCI	&  031\textunderscore S\textunderscore 0618    &  16598             &       15NC \\

033\textunderscore S\textunderscore 0724	 & 	19772	 & 	15AD & 037\textunderscore S\textunderscore  0566   &     16886          &       15MCI	& 010\textunderscore S\textunderscore 0420   &     17078          &       15NC \\

128\textunderscore S\textunderscore 0740	 & 	19990	 & 	15AD & 037\textunderscore S\textunderscore  0539  &     17018          &       15MCI	& 126 \textunderscore S\textunderscore  0506  &    17184           &       15NC \\

021\textunderscore S\textunderscore 0753	 & 	20169	 & 	15AD & 137\textunderscore S\textunderscore   0443  &     17030         &       15MCI	& 005\textunderscore S\textunderscore  0610  &     17303          &       15NC \\

137\textunderscore S\textunderscore 0796	 & 	23112	 & 	15AD & 005\textunderscore S\textunderscore  0546  &     17056          &       15MCI	&  006\textunderscore S\textunderscore  0484  &    17377           &       15NC \\

029\textunderscore S\textunderscore 0836	 & 	23231	 & 	15AD & 137\textunderscore S\textunderscore  0631  &    17109           &       15MCI	&  014\textunderscore S\textunderscore  0558  &     17400          &       15NC  \\

100\textunderscore S\textunderscore 0747	 & 	23581	 & 	15AD & 027\textunderscore S\textunderscore  0644  &     17157          &       15MCI	&  021\textunderscore S\textunderscore 0647   &    17668           &       15NC \\

127\textunderscore S\textunderscore 0754	 & 	23787	 & 	15AD & 133\textunderscore S\textunderscore  0629  &      17596         &       15MCI	& 137 \textunderscore S\textunderscore  0686  &  17813             &       15NC \\

012\textunderscore S\textunderscore 0803	 & 	24863	 & 	15AD & 021\textunderscore S\textunderscore  0626  &       17687       &       15MCI	&  032\textunderscore S\textunderscore 0677   &   17820            &       15NC \\

033\textunderscore S\textunderscore 0889	 & 	25026	 & 	15AD & 098\textunderscore S\textunderscore  0667  &       17702        &       15MCI	&  002\textunderscore S\textunderscore  0685  &  18211             &       15NC \\

126\textunderscore S\textunderscore 0891	 & 	25172	 & 	15AD & 052\textunderscore S\textunderscore  0671  &       17849        &       15MCI	 & 094\textunderscore S\textunderscore  0711  &   18589            &       15NC \\

005\textunderscore S\textunderscore 0929	 & 	25645	 & 	15AD & 014\textunderscore S\textunderscore  0563  &      17876         &       15MCI	& 127\textunderscore S\textunderscore 0684   &    18896           &       15NC \\

006\textunderscore S\textunderscore 0547	 & 	25816	 & 	15AD & 007\textunderscore S\textunderscore  0698  &     18363          &       15MCI	&  033\textunderscore S\textunderscore  0734  &     19155          &       15NC \\

002\textunderscore S\textunderscore 0955	 & 	26170	 & 	15AD & 133\textunderscore S\textunderscore  0638  &    18672           &       15MCI	 &  033\textunderscore S\textunderscore 0741   &   19258            &       15NC \\

130\textunderscore S\textunderscore 0956	 & 	27032	 & 	15AD & 033\textunderscore S\textunderscore  0723  &    19014           &       15MCI	&  094\textunderscore S\textunderscore  0692  &     19567          &       15NC \\

053\textunderscore S\textunderscore 1044	 & 	27782	 & 	15AD & 032\textunderscore S\textunderscore   0718 &    19035           &       15MCI	& 009 \textunderscore S\textunderscore  0751  &      20013         &       15NC \\

133\textunderscore S\textunderscore 1055	 & 	29381	 & 	15AD & 126\textunderscore S\textunderscore  0708  &     19089          &       15MCI	& 116\textunderscore S\textunderscore  0648  &     20370          &       15NC \\

100\textunderscore S\textunderscore 1062	 & 	29579	 & 	15AD & 128\textunderscore S\textunderscore  0715  &      19225         &       15MCI	&  129\textunderscore S\textunderscore 0778   &      20543         &       15NC \\

029\textunderscore S\textunderscore 1056	 & 	30618	 & 	15AD & 033\textunderscore S\textunderscore  0725 &    19404           &       15MCI & 029\textunderscore S\textunderscore 0824   &     23213          &       15NC	 \\ 

029\textunderscore S\textunderscore 0999	 & 	31239	 & 	15AD & 137\textunderscore S\textunderscore  0669  &     19419          &       15MCI	&116\textunderscore S\textunderscore  0657  &     23350          &       15NC \\

006\textunderscore S\textunderscore 0653	 & 	31252	 & 	15AD & 116\textunderscore S\textunderscore  0649  &   19516            &       15MCI	& 006\textunderscore S\textunderscore  0731  &    23468           &       15NC \\

014\textunderscore S\textunderscore 1095	 & 	31576	 & 	15AD & 130\textunderscore S\textunderscore  0505  &   19701            &       15MCI	& 029\textunderscore S\textunderscore  0845  &     24249          &       15NC  \\

094\textunderscore S\textunderscore 1090	 & 	31678	 & 	15AD & 137\textunderscore S\textunderscore  0722  &   19707            &       15MCI	& 009\textunderscore S\textunderscore  0862  &     25128          &       15NC \\

021\textunderscore S\textunderscore 1109	 & 	31784	 & 	15AD & 126\textunderscore S\textunderscore   0709 &   19754            &       15MCI  &098\textunderscore S\textunderscore  0896  &    25255           &       15NC \\

024\textunderscore S\textunderscore 1171	 & 	35190	 & 	15AD & 128\textunderscore S\textunderscore   0770 &   19907            &       15MCI	& 033\textunderscore S\textunderscore  0923  &    25427           &       15NC \\

133\textunderscore S\textunderscore 1170	 & 	35211	 & 	15AD & 014\textunderscore S\textunderscore  0658  &    20003           &       15MCI  & 130\textunderscore S\textunderscore  0886  &      25455         &       15NC \\

031\textunderscore S\textunderscore 1209	 & 	36178	 & 	15AD & 137\textunderscore S\textunderscore  0668  &    20202           &       15MCI	& 006\textunderscore S\textunderscore 0498   &   25790            &       15NC \\

130\textunderscore S\textunderscore 1201	 & 	36269	 & 	15AD & 137\textunderscore S\textunderscore   0800  &    20500           &       15MCI	 & 052\textunderscore S\textunderscore  0951  &   26642            &       15NC \\

027\textunderscore S\textunderscore 1081	 & 	37145	 & 	15AD & 002\textunderscore S\textunderscore  0782  &   20519            &       15MCI & 130\textunderscore S\textunderscore  0969  &    26688           &       15NC \\

126\textunderscore S\textunderscore 1221	 & 	37339	 & 	15AD & 130\textunderscore S\textunderscore 0783   &   20794            &       15MCI	&  021\textunderscore S\textunderscore  0984  &     27056          &       15NC \\ 

029\textunderscore S\textunderscore 1184	 & 	37350	 & 	15AD & 116\textunderscore S\textunderscore 0752   &  23097             &       15MCI	 &  024\textunderscore S\textunderscore  0985  &     27607          &       15NC \\

027\textunderscore S\textunderscore 1254	 & 	37859	 & 	15AD &  068\textunderscore S\textunderscore 0802   &  23389             &       15MCI	 &  024\textunderscore S\textunderscore  1063  &     28111          &       15NC \\

130\textunderscore S\textunderscore 1290	 & 	38395	 & 	15AD & 133\textunderscore S\textunderscore 0792   & 23444              &       15MCI	 & 033\textunderscore S\textunderscore  1098  &    30304           &       15NC \\

033\textunderscore S\textunderscore 1285	 & 	38593	 & 	15AD & 006\textunderscore S\textunderscore 0675   &  23644             &       15MCI	&  010\textunderscore S\textunderscore  0472  &     30481          &       15NC \\

033\textunderscore S\textunderscore 1283	 & 	38617	 & 	15AD & 031\textunderscore S\textunderscore 0821   &  23658             &       15MCI	&  137\textunderscore S\textunderscore 0972   &     31702          &       15NC \\

033\textunderscore S\textunderscore 1308	 & 	40114	 & 	15AD & 133\textunderscore S\textunderscore  0771  & 23876              &       15MCI	&  033\textunderscore S\textunderscore  1086  &     32054          &       15NC \\

024\textunderscore S\textunderscore 1307	 & 	41527	 & 	15AD & 133\textunderscore S\textunderscore  0727  &  23939             &       15MCI	 &130\textunderscore S\textunderscore  1200  &     36281          &       15NC \\

007\textunderscore S\textunderscore 1339	 & 	42344	 & 	15AD &  027\textunderscore S\textunderscore 0835    & 24138              &       15MCI &  116\textunderscore S\textunderscore 1232   &     37848          &       15NC  \\

130\textunderscore S\textunderscore 1337	 & 	42930	 & 	15AD & 031\textunderscore S\textunderscore  0830  &    24281          &       15MCI	 & 027\textunderscore S\textunderscore  0120  &   10933            &       15NC \\

127\textunderscore S\textunderscore 1382	 & 	45060	 & 	15AD & 029\textunderscore S\textunderscore 0878   &    24533           &       15MCI  & 068\textunderscore S\textunderscore  0127  &   11133            &       15NC \\

094\textunderscore S\textunderscore 1397	 & 	51790	 & 	15AD & 136\textunderscore S\textunderscore  0695  &     24585          &       15MCI	 &   068\textunderscore S\textunderscore 0210   &      11235         &       15NC \\

094\textunderscore S\textunderscore 1402	 & 	54220	 & 	15AD & 031\textunderscore S\textunderscore 0867   &   24962            &       15MCI	 &  136\textunderscore S\textunderscore  0186  &     11335          &       15NC \\

136\textunderscore S\textunderscore 0299        &      15181       &      30AD & 033\textunderscore S\textunderscore  0906  &  25053             &       15MCI & 009\textunderscore S\textunderscore 0842 & 24339 & 15NC  \\

136\textunderscore S\textunderscore 0426        &      16172       &      30AD & 033\textunderscore S\textunderscore  0922  &    25092            &       15MCI & 029\textunderscore S\textunderscore 0843& 24406 & 15NC \\

018\textunderscore S\textunderscore 0335        &      16560       &      30AD & 012\textunderscore S\textunderscore 0932   &    25150           &       15MCI &   032\textunderscore S\textunderscore 1169 & 34067 & 15NC\\

136\textunderscore S\textunderscore 0300        &      16719       &      30AD & 137\textunderscore S\textunderscore  0825  &    25272           &       15MCI & 018\textunderscore S\textunderscore 0055 &  9136 & 15NC\\

018\textunderscore S\textunderscore 0633        &      19093       &      30AD & 116\textunderscore S\textunderscore 0834   &    25467           &       15MCI & 100\textunderscore S\textunderscore   0015 &      8390         &       30NC  \\

012\textunderscore S\textunderscore 0689        &      19210       &      30AD &  094\textunderscore S\textunderscore  0921  &     25498          &       15MCI &    136\textunderscore S\textunderscore  0196  &     14236          &       30NC \\

126\textunderscore S\textunderscore 0606        &      20487       &      30AD & 136\textunderscore S\textunderscore  0873  &     25559          &       15MCI & 136\textunderscore S\textunderscore   0086 &     14712          &       30NC \\

131\textunderscore S\textunderscore 0691        &      20681       &      30AD & 100\textunderscore S\textunderscore  0930  &      25618         &       15MCI & 018\textunderscore S\textunderscore  0369  &       15110        &       30NC  \\

005\textunderscore S\textunderscore 0814        &      24734       &      30AD & 133 \textunderscore S\textunderscore  0912  &     26000          &       15MCI & 131\textunderscore S\textunderscore  0441  &     15959          &       30NC  \\

002\textunderscore S\textunderscore 0816        &      25405       &      30AD & 032\textunderscore S\textunderscore  0978  &    26407           &       15MCI & 032\textunderscore S\textunderscore  0479  &   16652            &       30NC \\

127\textunderscore S\textunderscore 0844        &      29230       &      30AD & 100\textunderscore S\textunderscore  0892  &     26443          &       15MCI &  018\textunderscore S\textunderscore 0425   &     17168          &       30NC \\

002\textunderscore S\textunderscore 1018        &      33832       &      30AD & 052\textunderscore S\textunderscore  0952  &    26661           &       15MCI &  126\textunderscore S\textunderscore 0405   &     17177          &       30NC\\

031\textunderscore S\textunderscore 4024        &    228879       &      30AD & 053\textunderscore S\textunderscore  0919  &   26739            &       15MCI & 005\textunderscore S\textunderscore  0553  &      17619         &       30NC \\

016\textunderscore S\textunderscore 4009        &    240946       &      30AD & 068\textunderscore S\textunderscore  0872  &    27450           &       15MCI & 126\textunderscore S\textunderscore  0605  &    17639           &       30NC \\

094\textunderscore S\textunderscore 4089        &    242719       &      30AD &  094\textunderscore S\textunderscore 1015   &    28005           &       15MCI & 005\textunderscore S\textunderscore 0602   &   19615            &       30NC\\

006\textunderscore S\textunderscore 4153        &    248517       &      30AD & 133\textunderscore S\textunderscore  1031  &    28152           &       15MCI &  012\textunderscore S\textunderscore  1009  &    28962           &       30NC \\

003\textunderscore S\textunderscore 4136        &    250173       &      30AD & 127\textunderscore S\textunderscore  0925  &   28165            &       15MCI &  012\textunderscore S\textunderscore  1212  &   37403            &       30NC\\

003\textunderscore S\textunderscore 4152        &    253760       &      30AD & 137\textunderscore S\textunderscore  0994  &    28269           &       15MCI & 007\textunderscore S\textunderscore 1206   & 37761              &       30NC   \\

098\textunderscore S\textunderscore 4215        &    255843       &      30AD & 009\textunderscore S\textunderscore  1030  &   28514            &       15MCI &068\textunderscore S\textunderscore 1191   &   38370            &       30NC \\

098\textunderscore S\textunderscore 4201        &    256178       &      30AD & 100\textunderscore S\textunderscore  0995  &    28877           &       15MCI &  007\textunderscore S\textunderscore  1222  &   38482            &       30NC\\

006\textunderscore S\textunderscore 4192        &    258594       &      30AD &  027\textunderscore S\textunderscore  1045  &   28947            &       15MCI &   094\textunderscore S\textunderscore 1241   &    41449           &       30NC \\

019\textunderscore S\textunderscore 4252        &    258947       &      30AD &  136\textunderscore S\textunderscore   0874 &     29140          &       15MCI & 002\textunderscore S\textunderscore  1261  &     41799          &       30NC \\

024\textunderscore S\textunderscore 4280        &    261332       &      30AD & 127\textunderscore S\textunderscore  1032  &     29177          &       15MCI &  002\textunderscore S\textunderscore 1280   &      41806         &       30NC \\

094\textunderscore S\textunderscore 4282        &    261855       &      30AD & 126\textunderscore S\textunderscore  0865  &      29243         &       15MCI &    052\textunderscore S\textunderscore  1251  &      43812         &       30NC \\

029\textunderscore S\textunderscore 4307        &    267595       &      30AD &  031\textunderscore S\textunderscore  1066  &   29388            &       15MCI &    100\textunderscore S\textunderscore 1286   &      45761         &       30NC \\

016\textunderscore S\textunderscore 4353        &    267937       &      30AD &  052\textunderscore S\textunderscore  0989  &     29525          &       15MCI &    094\textunderscore S\textunderscore 1267   &    46457           &       30NC \\

109\textunderscore S\textunderscore 4378        &    270669       &      30AD &  137\textunderscore S\textunderscore 0973   &    29650           &       15MCI &    131\textunderscore S\textunderscore  1301  &     49328          &       30NC \\

126\textunderscore S\textunderscore 4494        &    281605       &      30AD &  012\textunderscore S\textunderscore 1033   &   29964            &       15MCI &  098\textunderscore S\textunderscore  4003  &     224603          &       30NC \\

127\textunderscore S\textunderscore 4500        &    283515       &      30AD &  033\textunderscore S\textunderscore  1116  &   30317            &       15MCI & 098\textunderscore S\textunderscore  4018  &    228788           &       30NC \\

007\textunderscore S\textunderscore 4568        &    287472       &      30AD & 029\textunderscore S\textunderscore  1073  &   30359            &       15MCI & 031\textunderscore S\textunderscore  4021  &    229148           &       30NC  \\

006\textunderscore S\textunderscore 4546        &    287994       &      30AD & 029\textunderscore S\textunderscore  1038  &     30395          &       15MCI &  012\textunderscore S\textunderscore  4026  &     238532          &       30NC\\

130\textunderscore S\textunderscore 4589        &    291219       &      30AD &  052\textunderscore S\textunderscore  1054  &     30580         &       15MCI &   098\textunderscore S\textunderscore  4050  &     238615          &       30NC\\

016\textunderscore S\textunderscore 4591        &    292433       &      30AD & 037\textunderscore S\textunderscore  1078  &    30960           &       15MCI &016\textunderscore S\textunderscore 4097  &     243556          &       30NC   \\

016\textunderscore S\textunderscore 4583        &    294209       &      30AD & 010\textunderscore S\textunderscore  0422  &    31015           &       15MCI &  016\textunderscore S\textunderscore 4952   &    337793           &       30NC\\

014\textunderscore S\textunderscore  4615       &    294334       &      30AD &  012\textunderscore S\textunderscore  0917  &     31725          &       15MCI &  016\textunderscore S\textunderscore 4121   &      246002         &       30NC\\

130\textunderscore S\textunderscore 4641        &    295961       &      30AD & 006\textunderscore S\textunderscore  1130  &     31799          &       15MCI &  006\textunderscore S\textunderscore 4150   &    249403           &       30NC\\

130\textunderscore S\textunderscore 4660        &    300034       &      30AD & 126\textunderscore S\textunderscore 1077   &   31850            &       15MCI &  127\textunderscore S\textunderscore  4148  &    250137           &       30NC\\

019\textunderscore S\textunderscore 4549        &    300335       &      30AD &  037\textunderscore S\textunderscore   0588 &    32151           &       15MCI &   003\textunderscore S\textunderscore  4119  &    250894           &       30NC \\

126\textunderscore S\textunderscore 4686        &    300818       &      30AD &  052\textunderscore S\textunderscore 1168   &   32349            &       15MCI & 127\textunderscore S\textunderscore  4198  &      254320         &       30NC \\

005\textunderscore S\textunderscore 4707        &    304663       &      30AD &  010\textunderscore S\textunderscore 0904   &     32497          &       15MCI &  002\textunderscore S\textunderscore  4213  &     254582          &       30NC\\

021\textunderscore S\textunderscore 4718        &    304749       &      30AD & 002\textunderscore S\textunderscore  1155  &      33393         &       15MCI &  031\textunderscore S\textunderscore  4218  &    255978           &       30NC\\

018\textunderscore S\textunderscore 4733        &    306069       &      30AD & 029\textunderscore S\textunderscore  0871  &     33717          &       15MCI & 002\textunderscore S\textunderscore 4225   &    257270           &       30NC \\

130\textunderscore S\textunderscore 4730        &    306384       &      30AD &  127\textunderscore S\textunderscore  1140  &    33761           &       15MCI &   002\textunderscore S\textunderscore  4262  &   259653            &       30NC \\

137\textunderscore S\textunderscore 4756        &    307118       &      30AD &  029\textunderscore S\textunderscore  0914  &     33775          &       15MCI &  941\textunderscore S\textunderscore  4100  &   259781            &       30NC\\

027\textunderscore S\textunderscore 4801        &    314034       &      30AD &  100\textunderscore S\textunderscore  1154  &   34258            &       15MCI & 002\textunderscore S\textunderscore 4264   &    259796           &       30NC \\

027\textunderscore S\textunderscore 4802        &    317195       &      30AD & 094\textunderscore S\textunderscore  1188  &   34619            &       15MCI &  021\textunderscore S\textunderscore  4276  &    260047           &       30NC \\

006\textunderscore S\textunderscore 4867        &    322012       &      30AD & 012\textunderscore S\textunderscore  1165  &    35052           &       15MCI & 029\textunderscore S\textunderscore  4290  &    260425           &       30NC \\

016\textunderscore S\textunderscore 4887        &    325649       &      30AD & 133\textunderscore S\textunderscore  0913  &     35171          &       15MCI &  098\textunderscore S\textunderscore  4275   &   261459            &       30NC\\

007\textunderscore S\textunderscore 4911        &    328196       &      30AD &  012\textunderscore S\textunderscore 1175   &   35342            &       15MCI &  094\textunderscore S\textunderscore 4234   &    261531           &       30NC\\

021\textunderscore S\textunderscore 4924        &    331257       &      30AD & 126\textunderscore S\textunderscore  1187  &      36364         &       15MCI &   018\textunderscore S\textunderscore  4257  &     262076          &       30NC\\

137\textunderscore S\textunderscore 4756        &    332930       &      30AD & 009\textunderscore S\textunderscore 1199   &   36373            &       15MCI &  136\textunderscore S\textunderscore  4269  &    264215           &       30NC\\

127\textunderscore S\textunderscore 4940        &    335512       &      30AD & 029\textunderscore S\textunderscore   1215 &    37129           &       15MCI &  029\textunderscore S\textunderscore 4279   &   265980            &       30NC\\

027\textunderscore S\textunderscore 4938        &    336926       &      30AD &  116\textunderscore S\textunderscore   0890 &    37182           &       15MCI &  021\textunderscore S\textunderscore  4335  &   266174            &       30NC\\

027\textunderscore S\textunderscore 4962        &    338558       &      30AD &  100\textunderscore S\textunderscore 1226   &      37251         &       15MCI &    130\textunderscore S\textunderscore   4343 &     266217          &       30NC \\

130\textunderscore S\textunderscore 4982        &    341787       &      30AD &  005\textunderscore S\textunderscore   1224  &   37284            &       15MCI &   018\textunderscore S\textunderscore  4349  &     266625          &       30NC \\

130\textunderscore S\textunderscore 4984        &    342274       &      30AD & 037\textunderscore S\textunderscore  1225  &   37364            &       15MCI &  129\textunderscore S\textunderscore  4369  &   267405            &       30NC\\

130\textunderscore S\textunderscore 4971        &    342338       &      30AD & 029\textunderscore S\textunderscore 1218   &    37373            &       15MCI &    130\textunderscore S\textunderscore   4352 &     267711          &       30NC \\

127\textunderscore S\textunderscore 4992        &    342697       &      30AD & 027\textunderscore S\textunderscore   1213 &     37393          &       15MCI &  129\textunderscore S\textunderscore 4371   &     268462          &       30NC\\

019\textunderscore S\textunderscore 5012        &    343916       &      30AD &  127\textunderscore S\textunderscore  1210  &     38319          &       15MCI &    018\textunderscore S\textunderscore 4313   &       268930        &       30NC \\

019\textunderscore S\textunderscore 5019        &    345663       &      30AD & 116\textunderscore S\textunderscore 1243   &      38462         &       15MCI &  019\textunderscore S\textunderscore 4367   &      269273         &       30NC \\

002\textunderscore S\textunderscore 5018        &    346242       &      30AD & 033\textunderscore S\textunderscore   1309 &    38837           &       15MCI &  007\textunderscore S\textunderscore 4387   &     269929          &       30NC\\

127\textunderscore S\textunderscore 5028        &    346696       &      30AD & 027\textunderscore S\textunderscore  1277  &    39715           &       15MCI &  036\textunderscore S\textunderscore 4389   &     270462          &       30NC \\

130\textunderscore S\textunderscore 4997        &    347410       &      30AD & 129\textunderscore S\textunderscore  1246  &     40237          &       15MCI &  003\textunderscore S\textunderscore  4350  &    270999           &       30NC\\

005\textunderscore S\textunderscore 5038        &    351432       &      30AD & 129\textunderscore S\textunderscore  1204  &     40398          &       15MCI &  129\textunderscore S\textunderscore  4422  &    272184           &       30NC\\

127\textunderscore S\textunderscore 5056        &    353203       &      30AD &  033\textunderscore S\textunderscore   1284 &     40881          &       15MCI & 018\textunderscore S\textunderscore   4399 &    272231           &       30NC \\

127\textunderscore S\textunderscore 5058        &    354636       &      30AD & 033\textunderscore S\textunderscore 1279   &      40902         &       15MCI &  018\textunderscore S\textunderscore  4400  &    273504           &       30NC\\

007\textunderscore S\textunderscore 0128        &    10007         &    15MCI &  029\textunderscore S\textunderscore  1318  &        41062       &       15MCI &    021\textunderscore S\textunderscore 4421   &    273564           &       30NC \\

010\textunderscore S\textunderscore 0161        &    10077         &    15MCI &  116\textunderscore S\textunderscore 1271   &        41321       &       15MCI &  029\textunderscore S\textunderscore   4383 &   273993            &       30NC \\

021\textunderscore S\textunderscore 0141        &    10173         &    15MCI &  094\textunderscore S\textunderscore  1330  &       41491        &       15MCI & 003\textunderscore S\textunderscore  4441  &    277108           &       30NC  \\

127\textunderscore S\textunderscore 0112        &     10419        &    15MCI & 121\textunderscore S\textunderscore  1322  &      42188         &       15MCI &  136\textunderscore S\textunderscore  4433  &      278511         &       30NC\\

128\textunderscore S\textunderscore 0135        &     10431        &     15MCI & 094\textunderscore S\textunderscore  1314  &       42694        &       15MCI & 006\textunderscore S\textunderscore 4449   &    279470           &       30NC \\

128\textunderscore S\textunderscore 0138        &     10438        &     15MCI & 052 \textunderscore S\textunderscore  1352  &       42876        &       15MCI &    031\textunderscore S\textunderscore 4474   &      280369         &       30NC \\

098\textunderscore S\textunderscore 0160        &     10466        &     15MCI &  123\textunderscore S\textunderscore    1300 &     43214          &       15MCI &    007\textunderscore S\textunderscore 4488   &    281560           &       30NC \\

123\textunderscore S\textunderscore 0108        &     10738        &     15MCI & 121\textunderscore S\textunderscore  1350  &   44122            &       15MCI & 006\textunderscore S\textunderscore 4485   &     281882          &       30NC \\

037\textunderscore S\textunderscore 0150        &     10773       &     15MCI &  072\textunderscore S\textunderscore  1211  &      44137         &       15MCI &  010\textunderscore S\textunderscore  4345  &   282005            &       30NC \\

027\textunderscore S\textunderscore 0116        &      10783       &     15MCI &  116\textunderscore S\textunderscore  1315  &    44143           &       15MCI &    031\textunderscore S\textunderscore  4496  &    282638           &       30NC \\

128\textunderscore S\textunderscore  0188       &      10897       &     15MCI & 052\textunderscore S\textunderscore  1346  &     44515          &       15MCI &   098\textunderscore S\textunderscore  4506  &    282934           &       30NC \\

014\textunderscore S\textunderscore 0169        &      10987      &      15MCI & 027\textunderscore S\textunderscore 1387   &    44748           &       15MCI &   094\textunderscore S\textunderscore  4459   &     283445          &       30NC \\

021\textunderscore S\textunderscore 0178        &       10993     &      15MCI & 024\textunderscore S\textunderscore 1393   &       44887        &       15MCI &  094\textunderscore S\textunderscore 4460   &     283573          &       30NC\\

128\textunderscore S\textunderscore 0205        &       11011      &      15MCI &  132\textunderscore S\textunderscore 0987   &      45815         &       15MCI &  010\textunderscore S\textunderscore 4442   &    283915           &       30NC \\

128\textunderscore S\textunderscore 0200        &       11012      &      15MCI & 029\textunderscore S\textunderscore 1384   &        47455       &       15MCI &   007\textunderscore S\textunderscore 4516   &    284424           &      30NC \\

037\textunderscore S\textunderscore 0182        &       11121      &      15MCI & 072\textunderscore S\textunderscore  1380   &    49799           &       15MCI &  029\textunderscore S\textunderscore  4385   &     285589          &       30NC \\

137\textunderscore S\textunderscore 0158        &       11127      &      15MCI & 094\textunderscore S\textunderscore  1398  &    53551           &       15MCI & 094\textunderscore S\textunderscore  4503  &     286222          &       30NC \\

128\textunderscore S\textunderscore 0225        &       11179     &      15MCI &  024\textunderscore S\textunderscore 1400   &     53739          &       15MCI & 073\textunderscore S\textunderscore  4559  &    286553           &       30NC \\
 
136\textunderscore S\textunderscore 0107        &       11227     &      15MCI & 094\textunderscore S\textunderscore  1417  &    60175           &       15MCI & 021\textunderscore S\textunderscore  4558  &    287527           &       30NC \\

032\textunderscore S\textunderscore 0214        &       11280     &      15MCI & 127\textunderscore S\textunderscore 1419   &    61670           &       15MCI &  109\textunderscore S\textunderscore  4499  &      288999         &       30NC \\

005\textunderscore S\textunderscore 0222        &       11299     &       15MCI & 137\textunderscore S\textunderscore 1414   &     64472          &       15MCI &  100\textunderscore S\textunderscore 4469   &     289564          &       30NC \\

027\textunderscore S\textunderscore 0179        &       11348     &       15MCI &  127\textunderscore S\textunderscore  1427  &       69355        &       15MCI &    100\textunderscore S\textunderscore  4511  &         289653      &       30NC \\

021\textunderscore S\textunderscore 0231        &       11430     &       15MCI & 037\textunderscore S\textunderscore  1421  &      70885         &       15MCI &   012\textunderscore S\textunderscore  4545  &      290413         &       30NC \\

007\textunderscore S\textunderscore 0249        &        11544    &       15MCI & 137\textunderscore S\textunderscore 1426   &    72082           &       15MCI &   053\textunderscore S\textunderscore  4578  &     290814          &       30NC \\

098\textunderscore S\textunderscore 0269        &        11615    &        15MCI & 007\textunderscore S\textunderscore  0041  &   8177            &       15MCI &  127\textunderscore S\textunderscore  4604  &      291523        &       30NC \\

130\textunderscore S\textunderscore 0289        &        11850   &        15MCI &  123\textunderscore S\textunderscore  0050  &      8648         &       15MCI & 007\textunderscore S\textunderscore   4620 &     293938         &       30NC  \\

021\textunderscore S\textunderscore 0273        &        11942   &        15MCI &  100\textunderscore S\textunderscore   0006 &     8793          &       15MCI &  127\textunderscore S\textunderscore   4645 &    295590           &       30NC \\

007\textunderscore S\textunderscore 0293       &         11982   &        15MCI & 007\textunderscore S\textunderscore   0101 &     9602          &       15MCI &  002\textunderscore S\textunderscore 4270       &      260581      &        30NC \\

031\textunderscore S\textunderscore 0294      &          12065   &      15MCI & 123\textunderscore S\textunderscore  0106  &       10126        &       15NC &  013\textunderscore S\textunderscore   4579 &    296776           &       30NC \\

021\textunderscore S\textunderscore 0276      &          12092   &      15MCI &  100\textunderscore S\textunderscore  0035  &   8120            &       15NC &   013\textunderscore S\textunderscore  4580  &   296859            &       30NC \\

128\textunderscore S\textunderscore 0227      &          12119   &      15MCI &  100\textunderscore S\textunderscore 0047   &    8899           &       15NC &  012\textunderscore S\textunderscore  4642  &    296878           &       30NC \\

027\textunderscore S\textunderscore 0256      &          12250   &     15MCI &  010\textunderscore S\textunderscore 0067   &        9093       &       15NC &   012\textunderscore S\textunderscore   4643 &      297693         &       30NC\\

130\textunderscore S\textunderscore 0285     &           12424   &      15MCI & 018\textunderscore S\textunderscore  0043  &      9324         &       15NC &  029\textunderscore S\textunderscore  4585  &      298523         &       30NC\\

098\textunderscore S\textunderscore 0288     &           12654   &       15MCI & 100\textunderscore S\textunderscore  0069  &     9417          &       15NC &  013\textunderscore S\textunderscore  4616  &      300089         &       30NC\\

007\textunderscore S\textunderscore 0344     &           12697   &      15MCI & 032\textunderscore S\textunderscore  0095  &     9680          &       15NC &  029\textunderscore S\textunderscore  4652  &     300886          &       30NC\\

021\textunderscore S\textunderscore 0332     &           12862   &      15MCI & 123\textunderscore S\textunderscore 0072   &    9752           &       15NC & 137\textunderscore S\textunderscore 4632   &      301677         &       30NC \\

128\textunderscore S\textunderscore 0258    &            13085   &      15MCI & 007\textunderscore S\textunderscore 0070    &     10027          &       15NC &  094\textunderscore S\textunderscore  4649  &     302926          &       30NC \\

027\textunderscore S\textunderscore 0307    &           13281    &      15MCI & 131\textunderscore S\textunderscore   0123 &    10043           &       15NC &  016\textunderscore S\textunderscore   4638 &       305882        &       30NC \\

123\textunderscore S\textunderscore 0390   &             13315  &       15MCI & 027\textunderscore S\textunderscore 0118   &        11370       &       15NC & 013\textunderscore S\textunderscore  4731  &     308178          &       30NC \\

031\textunderscore S\textunderscore  0351  &            13783   &       15MCI & 098\textunderscore S\textunderscore 0172   &      11398   &       15NC & 136\textunderscore S\textunderscore  4726  &    308396           &       30NC \\

021\textunderscore S\textunderscore   0424 &           13909    &       15MCI &  130\textunderscore S\textunderscore   0232 &    11567           &       15NC  & 016\textunderscore S\textunderscore   4688 &    310327           &       30NC\\

053\textunderscore S\textunderscore  0389  &            13938   &       15MCI & 005\textunderscore S\textunderscore  0223  &     11645          &       15NC & 019\textunderscore S\textunderscore  4835  &  315857             &       30NC\\

094\textunderscore S\textunderscore  0434  &             13964  &       15MCI &  123\textunderscore S\textunderscore  0113  &     11714          &       15NC  & 127\textunderscore S\textunderscore 4843   &    316771           &       30NC\\

068\textunderscore S\textunderscore  0401  &            14161   &       15MCI &  128\textunderscore S\textunderscore  0230  &     11806          &       15NC  & 003\textunderscore S\textunderscore  4839  &   319414            &       30NC\\

131\textunderscore S\textunderscore  0409  &             14240  &       15MCI &  137\textunderscore S\textunderscore 0283   &    12028           &       15NC & 003\textunderscore S\textunderscore  4840  &     319427          &       30NC\\

116\textunderscore S\textunderscore  0361  &             14296  &       15MCI &  128\textunderscore S\textunderscore  0245  &    12242           &       15NC  & 003\textunderscore S\textunderscore  4872  &   321376            &       30NC\\

132\textunderscore S\textunderscore   0339 &              14367 &       15MCI &  128\textunderscore S\textunderscore 0272   &     12313          &       15NC & 003\textunderscore S\textunderscore  4900  &   325729            &       30NC \\

037\textunderscore S\textunderscore  0377  &             14405  &       15MCI & 128\textunderscore S\textunderscore 0229   &      12459         &       15NC & 016\textunderscore S\textunderscore 4951   &    337692           &       30NC \\

027\textunderscore S\textunderscore   0485 &             14928  &       15MCI & 021\textunderscore S\textunderscore  0337  &    12466           &       15NC & & & \\

130\textunderscore S\textunderscore   0102 &      9709         &       15MCI & 098\textunderscore S\textunderscore   0171 &     10818          &       15NC & & & \\

\bottomrule
\end{longtable}
\end{center}

\end{document}